\documentclass[manuscript]{aastex}

\shorttitle{Arrival time of CMEs using heliospheric imager observations}
\shortauthors{Mishra \& Srivastava}

\begin{document}

\title{Estimating arrival time of Earth-directed CMEs at \textit{in-situ} spacecraft using COR \& HI observations from STEREO}

\author{Wageesh Mishra\altaffilmark{1} and Nandita Srivastava\altaffilmark{1}}
\affil{Udaipur Solar Observatory, Physical Research Laboratory, P. O. Box 198, Badi Road,
Udaipur 313001, India}
\email{wageesh@prl.res.in}

\altaffiltext{1}{Udaipur Solar Observatory, Physical Research Laboratory, P. O. Box 198, Badi Road,
Udaipur 313001, India}

\begin{abstract}
The prediction of the arrival time and transit speed of CMEs near the Earth is one of the key problems in understanding the solar terrestrial relationship. Although, STEREO observations now provide a multiple view of CMEs in the heliosphere, the true speeds derived from stereoscopic reconstruction of SECCHI coronagraph data are not quite sufficient in accurate forecasting of arrival time of a majority of CMEs at the Earth. This is due to many factors which change the CME kinematics, like interaction of two or more CMEs or the interaction of CMEs with the pervading solar wind. In order to understand the propagation of CMEs, we have used the 3D triangulation method on SECCHI coronagraph (COR2) images, and geometric triangulation on the J-maps constructed from Heliospheric Imagers HI1 and HI2 images for eight Earth-directed CMEs observed during 2008-2010. Based on the reconstruction, and implementing the Drag Based Model for the distance where the CMEs could not be tracked unambiguously in the interplanetary medium, the arrival time of these CMEs have been estimated. These arrival times have also been compared with the actual arrival time as observed by \textit{in-situ} instruments. The analysis reveals the importance of heliospheric imaging for improved forecasting of the arrival time and direction of propagation of CMEs in the interplanetary medium.

\end{abstract}

\keywords{Coronal Mass Ejections - Solar Wind - Shock waves - Solar-terrestrial relations - Sun: heliosphere}

\section{Introduction} Coronal Mass Ejections (CMEs) are huge magnetized plasma eruptions from the Sun into interplanetary space which are observed in coronagraphic field of view as appearance and outward motion of new discrete, bright, white-light features over a time scale of tens of minutes \citep{Hundhausen1984}. CMEs are known as potential drivers of geomagnetic storms and many space weather effects \citep{Gosling1990, Richardson2001, Echer2008}. Heliospheric counterparts of CMEs are called Interplanetary Coronal Mass Ejections or ICMEs \citep{Zhao1992,Dryer1994}. The precise relation between a CME and ICME is not well understood because there is no unique signature to identify an ICME. Although, the term ICME was originally coined to distinguish between remote sensing observations of CMEs near the Sun by coronagraphs and their near-Earth observations by \textit{in-situ} spacecraft, in the present STEREO era, where CMEs can be tracked from near the Sun to the Earth and beyond, the term ICME has become redundant \citep{Webb2012}. Therefore, in this paper we use the term CME throughout for both, CME \& ICME.

Most of the studies which have been carried out to calculate travel time of CMEs from the Sun to the Earth, suffer from a lot of assumptions regarding geometry and evolution of a CME in interplanetary (IP) medium \citep{Howard2009,Vrsnak2010}. Earth directed CMEs having southward interplanetary magnetic field component interact with the magnetosphere at the day-side magnetopause. In this interaction, solar wind energy is transferred to the magnetosphere, primarily via magnetic reconnection which produces non-recurrent geomagnetic storms \citep{Dungey1961}. The rate of magnetic reconnection depends upon the magnitude of interplanetary convective electric field (V $\times$ B) \citep{Gosling1991}. Also, it has been shown that 83$\%$ of intense geomagnetic storms are due to CMEs \citep{Zhang2007}, therefore it is necessary to estimate the arrival time and transit speeds of CMEs near the Earth well in advance, to predict a majority of space weather events.

With the launch of \textit{SOlar and Heliospheric Observatory (SOHO)}, studies of evolution of CMEs were carried out that were primarily focused on observations by \textit{Large Angle and Spectrometric COronagraph} (LASCO; \citealt{Brueckner1995}) on-board SOHO and \textit{in-situ} identification of CMEs near the Earth by ACE \citep{stone1998} and WIND \citep{Ogilvie1995} combined with modeling efforts \citep{Wood1999, Andrews1999}. The understanding about kinematics of CMEs was based on two point measurements, one near the Sun up to a distance of 30 R$_\sun$ using coronagraph images and the other, near the Earth, using \textit{in-situ} instruments. Using the LASCO images, one could also estimate projected speeds of CMEs, although we lacked information about the true speed and direction of the Earth-directed CMEs.

An empirical CME arrival model (ECA) has been developed by \citet{Gopalswamy2001} in which they considered that a CME has an average acceleration up to a distance of 0.7 - 0.95 AU. After the acceleration cessation, a CME is assumed to move with a constant speed. They found that average acceleration has a linear relationship with initial plane-of-sky speed of the CME. ECA model has been able to predict arrival time of CMEs within an error of approximately $\pm$ 35 hours with an average error of 10.7 hours. Later, empirical shock arrival (ESA) model was developed which was able to predict the arrival time of CMEs within an error of approximately $\pm$ 30 hours with an average error of 12 hours \citep{Gopalswamy2005}. ESA model is a modified version of ECA model in which a CME is considered as the driver of MHD shocks. The other assumption is that fast mode MHD shocks are similar to gas dynamic shocks. Therefore, gas dynamic piston-shock relationship is utilized in ESA model. Various models have been developed to forecast the CME arrival time at 1 AU, based on empirical relationship between measured projected speeds and arrival time characteristics of various events (e.g., \citealt{Gopalswamy2001, Vrsnak2002, Schwenn2005}). Also, the analytical drag based model (e.g., \citealt{Vrsnak2007, Lara2009, Vrsnak2010}) as well as numerical MHD simulations models (e.g., \citealt{Odstrcil2004, Manchester2004, Smith2009}) have been developed. The efficacy of all these models to predict CME arrival time has been analyzed in various studies (e.g., \citealt{Dryer2004, Feng2009, Smith2009}). These studies show that the predicted arrival time is usually within an error of $\pm$ 10 hours but sometimes the errors can be larger than 24 hours. Many studies have also shown significant interaction of a CME with the ambient solar wind as they propagate in the interplanetary medium, resulting in acceleration of slow CMEs and deceleration of fast CMEs towards ambient solar wind speed \citep{Lindsay1999, Gopalswamy2000,Gopalswamy2001, Yashiro2004, Manoharan2006, Vrsnak2007}.

After the launch of twin \textit{Solar TErrestrial RElations Observatory} (STEREO; \citealt{Kaiser2008}) spacecraft, three dimensional (3D) aspects of CMEs could be studied for the first time using its \textit{Sun Earth Connection Coronal and Heliospheric Investigation} (SECCHI; \citealt{Howard2008}) coronagraph (COR) and Heliospheric Imager (HI) data. This is because the angular separation between the STEREO spacecraft and Sun-Earth line, enables to recognize the effect of two view directions on the coronagraph and heliospheric images which are always projections on the plane of sky. This led to the development of various 3D reconstruction techniques; (viz., tie-pointing: \citealt{Inhester2006}, forward modeling: \citealt{Thernisien2009}, polarization ratio: \citealt{Moran2004}). Also, several other methods which are derivatives of the tie-pointing method: 3D height-time technique (3D-HT) \citep{Mierla2008}, Local correlation tracking and triangulation (LCT-TP) \citep{Mierla2009}, triangulation of the center of mass (CM) \citep{Boursier2009} have been devised to estimate the true heliographic coordinates of CMEs in COR field of view (FOV). However, the estimation of true speeds of CMEs using 3D reconstruction method in coronagraphic (COR2) field of view is not often sufficient to predict accurately the arrival time near the Earth due to significant changes in their dynamics over different segments of journey in the IP medium. This may occur either due to varying ambient solar wind conditions or sometimes due to CME - CME interaction in the IP medium. The kinematics of CMEs over a range of heliocentric distances and their interaction in the IP medium has been investigated by exploiting the STEREO/HI observations, e.g. \citet{Harrison2012, Liu2012, Lugaz2012}. Also, \citet{Davis2009} showed that the apparent acceleration and deceleration noticed in the elongation variation of a CME from one STEREO view point can be used to obtain its kinematics and arrival time. \citet{Temmer2011} have implemented the ``corrected'' harmonic mean method to estimate the kinematics of CMEs and compared these with respect to the simulated background solar wind from the three dimensional MHD model. \citet{Byrne2010} have implemented the elliptical tie-pointing technique on COR and HI observations and quantified the angular width and deflecting trajectory of a CME. In their study, the derived kinematics was used as inputs in the ENLIL model to predict the arrival time of a CME at first Lagrangian point near the Earth. \citet{Maloney2010}, studied 3D kinematics of CMEs in the inner heliosphere using STEREO observations and pointed out different forms of drag force for fast and slow CMEs. The finding that CMEs near the Sun have a large range of speeds, typically varying between 100 to 2500 km s$^{-1}$ \citep{Yashiro2004} but a narrower range of speeds, typically varying between 300 to 1000 km s$^{-1}$  near 1 AU \citep{Gopalswamy2000,Schwenn2005,Richardson2010} also clearly underlines the influence of aerodynamic drag experienced by CMEs in interplanetary medium \citep{Cargill2004}. Each CME has different initial characteristics viz., initial speed, expansion speed, mass, angular width, density, magnetic field and is ejected in different solar wind conditions which coupled together, decide its dynamics in the interplanetary medium. Therefore, estimation of CME arrival time using any prediction model, with true speed estimated from 3D reconstruction method using coronagraph (COR2) data, as the only input, may not be accurate. Recent statistical study of CME kinematics based on estimation of 3D speed using Forward modeling \citep{Thernisien2009} and implementing \citet{Gopalswamy2000} empirical model, have reported an average error of nearly 10 hours between observed and predicted travel time \citep{Kilpua2012}.

Our present study of eight CMEs, attempts to understand the 3D propagation of CMEs by tracking them continuously throughout the interplanetary medium. For this purpose, we have used the geometric triangulation (referred to hereafter as GT) technique developed by \citet{Liu2010} on the time-elongation maps \citep{Sheeley1999}, popularly known as J-maps, constructed from COR2 and HI observations. Based on the GT using coronagraph (COR2) and Heliospheric Imager (HI) images, true kinematics of a CME is estimated. These estimated values are used as inputs in the Drag Based Model \citep{Vrsnak2012} beyond the distance where a CME could not be tracked unambiguously, and its arrival time as well as transit velocity at first Lagrangian (L1) point is predicted. The predicted arrival time and transit velocity of the CME at L1 is then compared with the actual arrival time and transit velocity as observed by \textit{in-situ} instruments e.g., ACE and WIND. Predicted arrival time is also compared with the arrival time estimated using true speed obtained by tie-pointing procedure (scc$\_$ measure: \citealt{Thompson2009}) on SECCHI/COR2 data alone.

In Section 2, we describe observations analysed in this paper and the implemented techniques of 3D reconstruction. Application of these techniques on eight CMEs is presented in Section 3. The obtained results and discussions are presented in Section 4 and a summary in Section 5. Our study reveals the importance of heliospheric imaging and GT technique on J-maps to improve the prediction of arrival time and transit velocity at L1 (near 1 AU) as well as direction of propagation of CMEs in the interplanetary medium.

\section{Observations and Analysis Techniques} Since its launch in 2006, the twin STEREO spacecraft (STEREO A $\&$ B)  has enabled  to image a CME in 3D from its birth in the corona to 1 AU and beyond. STEREO A moves faster and is slightly closer to the Sun than the Earth and leads the Earth in its orbit while STEREO B is little farther and trails the Earth. The separation between the two spacecraft increases by about 45$\arcdeg$ per year. Both spacecraft have identical imaging suite package named SECCHI. The STEREO/SECCHI is a suite of five telescopes (EUVI: 1-1.7 R$_\sun$, COR1: 1.5-4.0 R$_\sun$, COR2: 2.5-15.0 R$_\sun$, HI1: 15-90 R$_\sun$, and HI2: 70-330 R$_\sun$) which image the CMEs continuously from the Sun to the Earth and beyond. Both coronagraphs (COR1 and COR2) are pointed on the Sun while both Heliospheric Imagers (HI1 and HI2) are off-pointed from the Sun at solar elongation of 14$\arcdeg$ and 53.7$\arcdeg$ respectively \citep{Eyles2009}. HI1 and HI2 have wide field-of-view (FOV) of 20$\arcdeg$ and 70$\arcdeg$ and have their optical axis aligned in the ecliptic plane. With these imaging instruments, a CME can be observed from 0.4$\arcdeg$ to 88.7$\arcdeg$ in solar elongation angle.

We carried out the tie-pointing procedure of 3D reconstruction \citep{Thompson2009} which is based on the epipolar geometry \citep{Inhester2006}, to estimate the 3D kinematics of CMEs in the COR2 FOV. We estimated the true kinematics of all eight CMEs selected for our study, by implementing this technique on SECCHI/COR2 images.

In order to understand the propagation of CMEs in the interplanetary medium, we used the GT technique developed by \citet{Liu2010}, which assumes that same feature of a CME is observed from two different view points and the difference in measured elongation angles (angle of the feature with respect to the Sun-spacecraft line) for the tracked feature from STEREO A and B is entirely due to two viewing directions. Using imaging observations and Sun-centered coordinate system, the elongation angle of a moving feature can be calculated. The details of GT technique in ecliptic plane applicable for a feature propagating between the two spacecraft has been explained in \citet{Liu2010}. To estimate the propagation direction and radial distance of CME features, equations (1-4) of \citet{Liu2010} are used in our study.

Schematic diagram for the location of twin spacecraft and tracked feature is shown in Figure 1. \citet{Liu2010} did not take the effect of Thomson scattering and geometry of CME into account. In principle, the Thomson scattering physics and geometry of CMEs need to be considered. The total scattered intensity received by an observer at a certain location depends upon the radial and tangential component of the scattered radiation from the scattering source \citep{Howard2009}. The tangential component of the scattered radiation does not depend on the scattering angle but the radial component does. However, in our selection of events, where a CME is moving nearly towards the Sun-Earth line, both view directions (line-of-sight AP and BP as shown in Figure 1) will be nearly symmetrically located from the Sun-Earth line. Therefore, scattering angles ($\chi_{A}$ \& $\chi_{B}$) for both the observers will only be slightly different and the resulting difference in the received radial intensity by both the observers (STEREO A \& B) will be small. The approximation that both observers view the same part of CME, may not be true when Earth-directed CMEs are at a large distance from the Sun (for view directions AX and BY as shown in Figure 1) and also near the sun for very wide or for fast expanding CMEs. It is also rather unlikely that the same feature of CME will be tracked in each successive images, even from one observation point. Therefore, it seems reasonable to take the geometry of CME into account. But idealistic assumptions made for geometry results in new errors in the estimated kinematics.

In the present study, we reconstruct the CMEs using straightforward and easy procedure of GT on HI observations to relate remote sensing observations with \textit{in-situ} observations for practical application to space weather prediction scheme, excluding the geometry and complex treatment of Thomson scattering physics. Optimistically, taking aforementioned points, it can be mentioned succinctly that resulting errors by implementing GT technique, even after neglecting many real effects, will be minimized particularly for Earth-directed CMEs events.

\section{Selected Events} In the present study, we have selected eight Earth-directed CMEs observed on different dates after the launch of STEREO spacecraft. All these CMEs have been observed by twin STEREO spacecraft with different separation angles between them, from their birth in corona through the inner heliosphere by coronagraphs and heliospheric imagers respectively. These selected CMEs were observed on 12 December 2008, 07 February 2010, 12 February 2010, 14 March 2010, 03 April 2010, 08 April 2010, 10 October 2010 and 26 October 2010. These CMEs were also observed by \textit{in-situ} instruments. Remote sensing observational data of CMEs by twin STEREO is taken from UKSSDC (\url{http://www.ukssdc.ac.uk/solar/stereo/data.html}). \textit{In-situ} observations of CMEs were obtained from ACE and WIND spacecraft situated at first Lagrangian (L1) point, upstream from the Earth. We used the OMNI data with 1 minute time resolution for solar wind parameters, e.g. magnetic field, proton velocity, proton density, proton temperature and plasma beta. We also used latitude and longitude of magnetic field vector data with time resolution of 1 hour. Combined OMNI data were taken from NASA CDAWeb (\url{http://cdaweb.gsfc.nasa.gov}). We present our analysis for each event sequentially; the event of 12 December 2008 has been described with significant depth to explain the implemented techniques. Analysis for the other events has been carried out by adopting exactly the same methodology as explained for the event of 12 December 2008 and has been summarized briefly.

\subsection{12 December 2008 CME} 
\subsubsection{Remote sensing observations}
This CME was observed in SECCHI/COR1 Ahead at 04:35 UT in the NE quadrant and in SECCHI/COR1 Behind at 04:55 UT in NW quadrant. SOHO/LASCO observed this as a partial halo CME having an angular width of 184$\arcdeg$ with a linear speed of 203 km s$^{-1}$ (online LASCO CME catalog; \url{http://cdaw.gsfc.nasa.gov/CME$\_$list/}, see, \citet{Yashiro2004}). SOHO/LASCO could follow this CME up to 12 R$_\sun$ where its quadratic speed is 322 km s$^{-1}$. SOHO/LASCO catalog reported this accelerating CME as a poor event. The CME was associated with a filament eruption which started at 03:00 UT in the north-east quadrant observed in SECCHI/Extreme Ultraviolet Imager (EUVI) Ahead (A) 304 {\AA} images. Appearance of the CME in the SECCHI Ahead COR2, HI1 and HI2 is displayed in Figure 2.  

STEREO/HI detects photospheric light scattered from free electrons in K-corona and interplanetary dust around the Sun (F-corona) similar to coronagraph \citep{Billings1966, Tappin1987, Vourlidas2006, Howard2009a}. It also detects the light from the stars, planets within its field of view. The F-corona is stable on a time scale far longer than nominal image cadence of 40 and 120 minutes for HI1 and HI2 camera respectively. It has been often observed that when CMEs leave the coronagraphic FOV and enter the HI FOV, the Thomson scattered signal is too low to identify a particular feature in both set of images obtained by STEREO A and B. Therefore, for tracking a CME in interplanetary medium, the method of time elongation map (J-map) is used, which was initially developed by \citet{Sheeley1999} for LASCO images. \citet{Rouillard2009} implemented the same technique on HI data to reveal the outward motion of plasma blobs in interplanetary medium. A running difference technique \citep{Sheeley1997}, in which each image is presented after subtraction of previous image, gives image with dominant K-corona contribution. By taking image difference, the changes in electron density between consecutive images are highlighted which reveal density enhancement and depletion in these images. 

We constructed the J-maps along the ecliptic plane using long term background subtracted running differences images of COR2, HI1 and HI2 taken by STEREO A and B spacecraft. Prior to this, the HI image pair were aligned to remove the stellar contribution from the difference images which requires precise pointing information of HI instrument \citep{Davies2009}. For this purpose, we used the level 2 HI data which was corrected for cosmic ray, shutterless readout, saturation effect, flat field and instrument offsets from spacecraft pointing. Also, long term background was subtracted from these images. For the coronagraph data, we used the Level 0 data and processed it before taking the running difference. We calculated the elongation and position angle for each pixel of difference images and extracted a strip of constant position angle interval along the position angle of the Earth for COR2 and HI1 and HI2 images. The position angle tolerance considered for COR2 is 5$\arcdeg$ and for, HI1 and HI2, 2.5$\arcdeg$. We binned the pixel of the extracted strip over a specific elongation angle bin size, viz., 0.01$\arcdeg$ for COR2 and 0.075$\arcdeg$ for HI1 and HI2. Further, we took the resistant mean for each bin to represent the intensity at a corresponding elongation angle. Resistant mean for each elongation bin was scaled to reveal significant elongation bin. These scaled resistant mean were stacked as a function of time and elongation which is known as time elongation map (J-map) and shown in Figure 3 for the CME of 12 December 2008. In this J-map, the bright curve having positive inclination reveals the evolution of CME. In the J-map constructed from images taken by STEREO A spacecraft, there are two nearly horizontal lines in HI2 FOV which start at elongation angle of 30.7$\arcdeg$ and 70.1$\arcdeg$ respectively. These lines are due to appearance of planets Venus and Earth respectively in HI2 A FOV. In the HI2 A images taken on 12 December 2008 at 00:09:21 UT, Venus is seen at position angle (helioprojective radial longitude) of 88.1$\arcdeg$ and elongation angle (helioprojective radial latitude) of 30.7$\arcdeg$ which corresponds to the pixels 830 $\times$ 487 in the image of 1024 $\times$ 1024 size. At this time, the Earth is observed corresponding to pixels 277 $\times$ 509 in the HI2 A image and to pixels 655 $\times$ 510 in the HI2 B image.  The appearance of planets in HI FOV saturates the pixels where they are found and also their signal bleeds up and down in the CCD which creates vertical columns of saturated pixels in the HI images, shown in Figure 2. In the J-map constructed from STEREO B images, one horizontal line is observed which starts at elongation angle of 63.9$\arcdeg$ and is due to appearance of planet Earth in HI2 B FOV, which results in vertical columns of saturated pixels in HI2 B images.

By tracking a CME continuously in the J-maps, separately for STEREO A and B images, independent elongation angles of a moving CME feature is estimated. Appropriate separation angle between STEREO A and B, their heliocentric distances and estimated elongation angles are used as inputs in GT technique \citep{Liu2010} to obtain the distance and propagation direction of the moving CME feature. Its velocity is calculated from estimated distance profile by using numerical differentiation with three-point Lagrange interpolation method. Figure 4 shows the kinematics of CME of 12 December 2008. In Figure 4, red vertical lines show the errors bars which are mathematically calculated, taking into account the uncertainty in measurements of elongation angles. We have considered uncertainty of 5 pixels in measurements of elongation angles which is equivalent to uncertainties of 0.02$\arcdeg$, 0.1$\arcdeg$ and 0.35$\arcdeg$ in elongation for COR2, HI1 and HI2 images respectively.

To estimate the true kinematics of 12 December 2008 CME in COR2 (2.5-15 R$_\sun$) field of view, we carried out 3D reconstruction of its selected features. For this, we used SECCHI/COR2 images and after calibration and background subtraction, we applied the tie-pointing method (scc$\_$measure: \citealt{Thompson2009}) on both set of images taken by COR2 A and B. On identification of a feature along the leading edge observed in both set of images, we found a correspondence between pixels of both images. The lines-of-sight corresponding to those pixels were found and tracking of these lines-of-sight backward; we obtained a point of intersection, which is the true 3D coordinate of selected feature \citep{Inhester2006}. The kinematics obtained after 3-D reconstruction using COR2 data is shown in Figure 5. The true stonyhurst heliographic latitude and longitude has been plotted in Figure 5  which shows that the CME was Earth-directed.

\subsubsection{Estimation of Arrival Time and Transit Speed at L1} Using GT method on HI data we have estimated the kinematics of CME of 12 December 2008 up to approximately 138 R$_\sun$. We assume that after traversing such a large distance, the speed of the CME will depend solely on the aerodynamic drag. Therefore, we can use quadratic form of drag acceleration  a = -$\gamma$ $(v-w)$ $|(v-w)|$, where $v$ is the instantaneous CME speed and $w$ is the ambient solar wind speed (\citealt{Cargill2004} and references therein). The drag parameter is expressed as $\gamma$ = $\frac{c_{d}A \rho_{w}}{M + M_{v}}$, where $c_{d}$ is the dimensionless drag coefficient, $A$ is the cross-sectional area of CME perpendicular to its direction of propagation, $\rho_{w}$ is the ambient solar wind density, $M$ is the mass of CME and $M_{v}$ is the virtual mass which is written as, $M_{v}$ = $\rho_{w} V/2$, where $V$ is the volume of CME.      

We used the drag based model (DBM) developed by \citet{Vrsnak2012} to estimate the arrival time of 12 December 2008 CME. In the DBM, it is assumed that after 20 R$_\sun$, drag parameter and ambient solar wind speed do not change with heliocentric distance. The DBM also assumes that, $A$ $\propto$ $r {^2}$, $\rho_{w}$ $ \propto$ $ r^{-2}$, $M$ = constant and $c_{d}$ = constant; and are valid after traversing a long distance. Moreover, virtual mass $M_{v}$ is assumed as negligibly small and with continuity equation for isotropic flow added with $\rho_{w}$ $ \propto$ $ r^{-2}$ gives $w$ = constant. By estimating the mass of the CME and considering it constant beyond distance of 20 R$_\sun$, solar wind density using \citet{Leblanc1998}, volume using cone model of  CME \citep{Xie2004} and the drag parameter can be calculated at any heliocentric distance. However, the statistical analysis of \citet{Vrsnak2012} for a sample of events shows that drag parameter lies in the range of 0.2 $\times$ 10$^{-7}$ - 2.0 $\times$ 10$^{-7}$ km$^{-1}$. Their statistical analysis shows that the ambient solar wind speed should be chosen to lie between 300 - 400 km s$^{-1}$ for slow solar wind environment. However, if there is a coronal hole in the vicinity of source region of CME, solar wind speed value should be chosen between 500 - 600 km s$^{-1}$. Since there is limiting accuracy and difficulty in the reliable estimation of variables (mass, cone angle of CME and solar wind density) on which the drag parameter depends, we used the extreme values of the range of     drag parameter for CMEs in our study. The limitation on accurate and reliable estimation of CME mass using single coronagraph images has been discussed in \citet{Vourlidas2000} and using multiple view point (SECCHI/COR) in \citet{Colaninno2009}. Choosing extreme values of the range of drag parameter, we estimate the maximum possible errors which can occur due to CMEs having different characteristics.

We used the estimated kinematics as inputs for three different approaches in order to predict the arrival time of CMEs at L1 ($\approx$ 1 AU). These three approaches are described as follows:\\
\textbf{(1) From DBM with inputs from GT on COR2 and HI:} We used the DBM, combined with the inputs obtained by implementing the GT technique on COR2 and HI observations. For the inputs in the DBM, the initial radial distance and CME take-off date and time is obtained from the last data points estimated in GT scheme. The initial take-off velocity is taken as the average of last few velocity points of fitted polynomial for estimated velocity in ecliptic plane. In this particular event of 12 December 2008, we used the DBM, with take-off date and time of CME as 15 December 2008 at 01:00 UT, take-off distance as 138.3 R$_\sun$, take-off velocity as 330 km s$^{-1}$, ambient solar wind speed as 350 km s$^{-1}$ and drag parameter 0.2 $\times$ 10$^{-7}$. Using these values, we obtained its arrival time as 20:25 UT on 16 December 2008 and transit speed 331 km s$^{-1}$ at L1 point. Using the maximum value of average range of drag parameter as 2 $\times$ 10$^{-7}$ and keeping other parameters same in the DBM, the predicted arrival time of CME was found to be at 19:55 UT on 16 December 2008 with transit speed of 338 km s$^{-1}$.\\
\textbf{(2) Using polynomial fit of distance estimated from GT on COR2 and HI:} The radial distance of a moving feature which is estimated by implementing GT technique is fitted to a second order polynomial and by extrapolation, we obtained the arrival time of CME at L1. The predicted arrival time is at 06:23 UT on 17 December 2008. In an earlier study by \citet{Liu2010} the predicted arrival time for this CME, by extrapolating the estimated distances is found to be during 16:00 - 18:00 UT on 16 December 2008. The difference in arrival time may arise due to different extrapolation techniques but is most likely due to tracking of different features in J-maps. It highlights the possible error in the arrival time prediction due to manual tracking of ambiguous edges of bright features in the J-maps. Since, fitting of estimated distance in second order polynomial includes all the points with different velocity phases together, it may result in greater uncertainty in extrapolated arrival time.\\
\textbf{(3) Using true speed estimated in COR2 FOV:} From the 3D reconstruction of CME leading edge using tiepointing (scc$\_$measure) method on SECCHI/COR2 data, the true velocity was found to be 453 km s$^{-1}$ at 14.4 R$_\sun$ at 13:07 UT on 12 December 2008. With a simple assumption that the speed of a CME is constant beyond COR2 FOV, the predicted arrival time is at 01:45 UT on 16 December 2008 at L1. 

In depth analysis of this CME has been carried out earlier by several researchers using STEREO COR and HI observations. \citet{Byrne2010} predicted the arrival time of this CME using 3D reconstruction of CME front and ENLIL simulation. They found that the predicted arrival time of CME agreed well with the CME front plasma pileup ahead of magnetic cloud observed \textit{in-situ} by ACE and WIND spacecraft. In the study of \citet{Davis2009}, three different features within the CME were tracked in the heliosphere from independent location of STEREO spacecraft and their speed, direction and arrival time at L1 point were estimated.

\subsubsection{\textit{In-situ} Measurements of CME} Near the Earth, CMEs can be identified by a combination of various signatures observed \textit{in-situ} \citep{Zurbuchen2006}. Figure 6 shows, the predicted arrival time and transit velocity at \textit{in-situ} spacecraft situated at L1 point corresponding to extreme values of average range of drag parameter used in DBM as discussed in previous section. The first  dashed vertical line (red) marks the arrival of sheath at 11:55 UT on 16 December and second dashed vertical line (red) marks the arrival of leading edge of magnetic cloud \citep{Klein1982} at 04:39 UT on 17 December. The third dashed vertical line (red) marks the trailing edge of magnetic cloud at 15:48 UT on 17 December 2008. The hatched region (blue) shows the predicted arrival time (with uncertainties due to range of values of drag parameter adopted in DBM) using drag based model. In the fourth panel, two red horizontal lines mark the predicted transit velocities at L1 of the tracked feature corresponding to different values of drag parameter used in the DBM. In the third panel (from the top), the red curve shows expected proton temperature calculated from observed \textit{in-situ} proton speed \citep{Lopez1986, Lopez1987}.     

\subsection{07 February 2010 CME} The SOHO/LASCO observed this CME on 07 February 2010 at 03:54 UT as a halo CME with a linear speed of 421 km s$^{-1}$. The speed of CME was nearly constant in the LASCO FOV. It appeared in the COR1 FOV of STEREO A and B at east and west limb respectively at 02:45 UT. We constructed the J-map for this CME as described for the 12 December 2008 CME (Section 3.1.1). At the time of this CME, the distance of STEREO A and B from the Sun was 0.96 and 1.01 AU respectively and separation angle between them was 135.6$\arcdeg$. In HI2 A and HI2 B FOV, the planets Earth and Mars were visible which caused the appearance of vertical column of saturated pixels in the images and parallel lines in the J-maps corresponding to their elongation angles. This event was tracked in the heliosphere and independent elongation of a moving feature of CME from two vantage points were estimated using J-maps. These estimated independent elongation angles and the positional inputs of twin STEREO are used in the GT scheme to estimate the kinematics of CME. The obtained kinematics is shown in Figure 7 in which a gap of estimated parameters is due to existence of singularity in GT scheme. The estimated points in this range have non-physical variations, therefore have been removed \citep{Liu2011}. We found that singularity occurs when summation of both independent elongation angles measured from two spacecraft and separation angle between two spacecraft becomes nearly 180$\arcdeg$. In this situation, the line-of-sight from two view points of spacecraft coincide along their entire length. Therefore, a single value of longitude can not be obtained in this scenario. In Figure 1, it is shown that line-of-sight from two locations of STEREO A and B (AP1 and BP1) will be parallel for the point P1 to be triangulated. Therefore, in this case, singularity will occur in triangulation scheme. As the STEREO mission progressed, separation angle between the two spacecraft increased and therefore, the occurrence of  singularity will be  an issue at lesser elongation angles.

We used the drag model with inputs from the last data point of estimated kinematics to calculate the arrival time of CME at L1 point. Using take-off distance as 171 R$_\sun$, take-off time as 10 February 2010 at 03:30 UT, take-off velocity as 455 km s$^{-1}$, drag parameter 0.2 $\times$ 10$^{-7}$ km$^{-1}$ and ambient solar wind speed as 350 km s$^{-1}$; we obtained the predicted arrival time at 21:40 UT on 10 February and transit speed 442 km s$^{-1}$ at L1 point. Keeping all the input parameters same but drag parameter as 2.0 $\times$ 10$^{-7}$ km$^{-1}$; the predicted arrival time is 22:50 UT on 10 February and transit speed 393 km s$^{-1}$. Using second order polynomial fit for estimated distance points and extrapolating it, we obtained the arrival time as 00:50 UT on 11 February at L1 point. We also carried out the 3D reconstruction of this CME using SECCHI/COR2 data and estimated the true heliographic coordinates of a selected feature along the leading edge. The true velocity was estimated as 480 km s$^{-1}$ at 13.5 R$_\sun$ at 06:39 UT on 07 February 2010. Assuming that the true velocity of CME is constant beyond COR2 FOV, the predicted arrival time is 14:55 UT on 10 February 2010 at L1.

The \textit{in-situ} observations of this CME is shown in Figure 8. The first vertical dashed line (red) at 01:00 UT on 11 February marks the arrival of a weak shock or CME sheath, second vertical dashed line (red) at 12:47 UT marks the arrival of CME leading edge and third vertical dashed line (red) at 23:13 UT marks the trailing edge of CME. The hatched region (blue) shows the predicted arrival time with uncertainties due to extreme values of drag parameter with the same inputs from estimated kinematics employed in the DBM.

\subsection{12 February 2010 CME} This CME was observed in NE quadrant in SECCHI/COR1 A observations and in NW quadrant in SECCHI/COR1 B images at 11:50 UT on 12 February. It was also observed by SOHO/LASCO at 13:42 UT as a halo CME with a linear speed of 509 km s$^{-1}$. It was observed to decelerate in LASCO FOV and its speed at final height ($\approx$ 25 R{$_\sun$}) was measured to be  358 km s$^{-1}$. We constructed the J-maps using COR2, HI1 and HI2 observations for the CME. In the SECCHI/HI2 A \& B FOV, planets Earth and Mars were visible as parallel lines corresponding to their elongation angles in J-maps. Independent elongation angle of moving feature from two view points was estimated by tracking the bright positively inclined feature in the J-maps corresponding to this CME. Implementing the GT technique, the kinematics of CME was estimated and is shown in Figure 8. There occurs a data gap in this plot which is due to occurrence of singularity in GT scheme, explained in details in Section 3.2.

To predict the arrival time of this CME, we combined the estimated kinematics with the DBM. Inputs parameters to be used in the DBM were calculated in the same way as explained earlier in Section 3.1.2. Using the inputs, e.g. initial take-off distance as 183 R$_\sun$, take-off time as 15 February 2010 at 01:44, take-off velocity $\sim$ 450 km s$^{-1}$, ambient solar wind speed as 350 km s$^{-1}$ and drag parameter 0.2 $\times$ 10$^{-7}$ km$^{-1}$ in the DBM; the predicted arrival time and transit speed of CME at L1 is 14:35 UT on 15 February 2010 and 442 km s$^{-1}$, respectively. If the drag parameter is taken as 2.0 $\times$ 10$^{-7}$ km$^{-1}$ in the DBM, keeping the rest of the input parameters same, the predicted arrival time of the CME is found to be at 15:20 UT on 15 February 2010 and transit speed 401 km s$^{-1}$.

The identification of the CME near the Earth is studied by analyzing the \textit{in-situ} data and shown in Figure 10. In this figure, the first vertical dashed line (red) at 18:42 UT on 15 February marks the arrival of shock, second dashed vertical line (red) at 04:32 UT on 16 February marks the arrival of CME leading edge and third vertical dashed line (red) at 12:38 UT on 16 February marks the CME trailing edge. The hatched region with blue color marks the predicted arrival time (with uncertainties due to extreme values of drag parameter used in the DBM) by using estimated kinematics parameters combined with DMB. Using second order polynomial fit for distance, we obtained the predicted arrival time of CME at L1 at 16:07 UT on 15 February. Further, from 3D reconstruction in COR2 FOV, the true velocity of leading edge of CME was estimated as 867 km s$^{-1}$ at a distance of 14.8 R$_\sun$ at 14:54 UT on 12 February. Considering that the CME speed was constant up to L1, the predicted arrival time at L1 is at 11:02 UT on 14 February.

\subsection{14 March 2010 CME} This CME was observed on 14 March 2010 by SECCHI/COR1 A in the NE quadrant and by SECCHI/COR1 B in NW quadrant of coronagraphic images. SOHO/LASCO observed this as a partial halo (angular width $\approx$ 260$\arcdeg$) CME at 00:30 UT on 14 March with a linear speed of 351 km s$^{-1}$. In the LASCO FOV, nearly constant speed of CME was observed. In the J-maps constructed from COR \& HI A images, this CME could be tracked nearly up to 35$\arcdeg$ while in J-maps constructed from COR \& HI  B images, tracking could be possible up to 50$\arcdeg$. In the SECCHI/HI2 A FOV, planets Earth and Mars were seen while in HI2 B FOV only Earth could be seen. We tracked the CME in the heliosphere and estimated its independent elongation from two STEREO locations. These elongation angles and separation angle between twin STEREO spacecraft were used as inputs in the GT scheme to obtain the propagation direction and distance of CME. Velocity is calculated from the adjacent distances using three point Lagrange interpolation. The kinematics of this CME is shown in Figure 11. The occurrence of singularity have been noticed for this case also, in the GT scheme and therefore, estimated kinematics in the time range of singularity, is not shown.

The estimated kinematics is used as inputs in the DBM to predict the arrival time of CME at L1 point.  We used the DBM with take-off distance as 135.9 R$_\sun$, take-off velocity as 460 km s$^{-1}$, take-off date and time as 16 March 2010 at 12:07 UT, ambient solar wind speed as 350 km s$^{-1}$ and drag parameter 0.2 $\times$ 10$^{-7}$ km$^{-1}$ as inputs and obtained its predicted arrival time at 21:10 UT on 17 March with transit velocity 437 km s$^{-1}$ at L1. Keeping all these inputs parameters same and using the maximum value of statistical range of drag parameter 2 $\times$ 10$^{-7}$ km$^{-1}$ in the DBM, we obtained the predicted arrival time as 01:00 UT on 18 March and transit velocity 378 km s$^{-1}$ at L1. Predicted arrival time of CME at L1, using second order polynomial fit for distance is, at 16:21 UT on 17 March. We also implemented the tie-pointing method of 3D reconstruction on the leading edge of the CME in the COR2 FOV and estimated the true kinematics of a CME feature. Assuming that the true speed (335 km s$^{-1}$) estimated at true height (11 R$_\sun$) at 03:54 UT on 14 March is constant beyond the COR2 FOV, the predicted arrival time of CME is at 00:17 UT on 19 March.

The \textit{in-situ} observations for this CME have been shown in Figure 12. In this figure, the first vertical dashed line (red) at 21:19 UT on 17 March marks the arrival of CME leading edge and second vertical dashed line (red) at 11:26 UT on 18 March marks the CME trailing edge. The hatched region (blue) shows the predicted arrival time of CME with range of uncertainty due to extreme values of the range of drag parameter employed in the DBM.

\subsection{03 April 2010 CME} This CME was detected by SOHO/LASCO on 03 April at 10:33 UT as a halo. It had a projected plane-of-sky linear speed of around 668 km s$^{-1}$ as measured from the LASCO images. It was observed at 09:05 UT by SECCHI/COR1 A in SE quadrant and by SECCHI/COR1 B in SE quadrant. The source region of the CME was NOAA AR 1059. The CME was accompanied by a filament disappearance, a post eruption arcade, coronal dimming, and EIT wave and B7.4 long duration flare peaking at 09:54 UT \citep{Liu2011}.

We constructed the J-maps to track the CME in the heliosphere. Due to appearance of Milky Way galaxy in the SECCHI/HI2 B images, the signal of this CME is not well pronounced, therefore, it could not be tracked beyond 27$\arcdeg$ elongation in the J-map constructed from STEREO B images. Planets Earth and Mars are visible in HI2 A images at 58.1$\arcdeg$ and 50.6$\arcdeg$. Earth is visible in the HI2 B images at 54.3$\arcdeg$ elongation. Independent elongation angles are extracted from the leading edge of the track of the CME in the J-maps. Then GT scheme is implemented to estimate the distance and propagation direction of CME. The estimated kinematics is displayed in Figure 13 and data gap of approximately 6 hours in the estimated kinematics is due to occurrence of singularity in implemented triangulation scheme.

We used the distance and velocity of CME at last estimated data point as inputs in the DBM to predict its arrival time at L1.  At the time of this CME, the Earth was blown over by high speed solar wind stream emanating from the northern polar coronal hole. Due to presence of this coronal hole and generally large spatial scale of CME its kinematics will be partly governed by the high speed stream as shown by \citet{Vrsnak2012}. They also showed that ambient solar wind speed of 550 Km s$^{-1}$ and low value of drag parameter should be considered as inputs in the DBM. We used the CME take-off speed as 640 km s$^{-1}$, take-off date and time as 04 April at 07:23 UT, take-off distance as 100.8 R$_\sun$ and drag parameter 0.2 $\times$ 10$^{-7}$ km$^{-1}$ in the DBM as inputs. We obtained the predicted arrival time of CME at 17:35 UT on 05 April and transit speed of 624 km s$^{-1}$ at L1 point. The predicted arrival time using extrapolation of second order polynomial fit of distance, is obtained as 09:00 UT  on 05 April. This CME has been studied in details by \citet{Liu2011}. They predicted the arrival by extrapolating the estimated distance at 12:00 UT on 05 April which is approximately the same as predicted by us using polynomial fit, with an error of 3 hours. We also carried out 3D reconstruction of CME leading edge using SECCHI/COR2 data and obtained its true kinematics. Assuming that this true estimated velocity (816 km s$^{-1}$) at 12:24 UT on 03 April is constant beyond COR2 FOV, the predicted arrival time of CME at L1 is 11:25 UT on 05 April.

We identified the CME \textit{in-situ} and the plot is shown in Figure 14. Here, first vertical dashed line (red) marks the arrival of shock at 8:28 UT on 05 April, second vertical dashed line (red) marks the arrival of CME leading edge at 13:43 UT and fourth vertical dashed line (red) marks the CME trailing edge at 16:05 UT on 06 April. Third vertical dashed line (blue) marks the predicted arrival time of CME obtained after the estimated dynamics employed in the DBM. In the fourth panel from the top, horizontal line (red) marks the predicted transit velocity of CME at L1. 

\subsection{08 April 2010 CME} In the SOHO/LASCO observations, this CME was detected at 04:54 UT on 08 April as a partial halo. The plane-of-sky speed of this CME was 264 km s$^{-1}$ which decelerated in LASCO FOV. SECCHI/COR1 A observed this CME in NE quadrant and COR1 B observed in NW quadrant at 03:25 UT on 08 April. The CME was accompanied by a B3.7 flare in the NOAA AR 11060.

Time-elongation plot (J-map) in the ecliptic plane was constructed for this CME. The CME was tracked in the heliosphere up to 54$\arcdeg$ elongation angle in J-maps constructed from SECCHI A observations while in J-maps constructed from SECCHI B observations it could be tracked up to 44$\arcdeg$ only. For comparison, planets Earth and Mars are seen at 58$\arcdeg$ and 48.4$\arcdeg$ elongation respectively, in HI2 A FOV on 08 April. In HI2 B FOV at this time, Earth is seen at 54.5$\arcdeg$ elongation. GT scheme is implemented to estimate the distance and propagation direction of CME in heliosphere. Estimated propagation direction of CME is converted to an angle with respect to the Sun-Earth line in the ecliptic plane. Its positive value implies that CME is moving in the west direction from the Sun-Earth line while negative value means CME is propagating in the east direction form the Sun-Earth line. The obtained kinematics is shown in Figure 15 and gap in the estimated kinematics for nearly 12 hours is due to occurrence of singularity in triangulation scheme.

We used the drag model with the inputs of take-off velocity as 550 km s$^{-1}$, take-off distance as 167.3 R$_\sun$, take-off date and time as 10 April at 17:07 UT, ambient solar wind speed as 350 km s$^{-1}$ and drag parameter 0.2 $\times$ 10$^{-7}$ km$^{-1}$. Using these inputs in the DBM, the predicted arrival time of CME is at 09:45 UT on 11 April and transit velocity 511 km s$^{-1}$ at L1. Keeping all the inputs parameters same and taking drag parameter value 2 $\times$ 10$^{-7}$ km$^{-1}$ in the DBM, the predicted arrival time of CME is at 12:55 UT on 11 April with transit speed 402 km s$^{-1}$. By extrapolating the second order polynomial fit of distance, the predicted arrival time of CME at L1 is at 06:32 UT on 11 April. On 3D reconstruction of the CME by using the tie-pointing technique on SECCHI/COR2 data, the true velocity of CME leading edge at true height of 12 R$_\sun$ was found to be 478 km s$^{-1}$ at 07:24 UT on 08 April. Assuming that CME speed is constant beyond COR2 FOV, its predicted arrival time at L1 is at 16:35 UT on 11 April.

By analyzing the \textit{in-situ} data taken nearly at 1AU, identification of CME boundary is shown in Figure 16. In this figure, a weak shock or sheath is marked by the first dashed vertical line (red) at 12:44 UT on 11 April. Leading and trailing edge of a magnetic cloud is marked by the second and third vertical dashed line (red) at 02:10 UT and 13:52 UT on 12 April, respectively. Hatched region with blue color shows the predicted arrival time of CME (with uncertainties) obtained using DBM.

\subsection{10 October 2010 CME} This CME was accompanied with a filament eruption in south east quadrant of the solar disc. LASCO on-board SOHO observed this event at 22:12 UT on 10 October as a slow (projected linear speed $\approx$ 262 km s$^{-1}$) and partial halo (angular width $\approx$ 150$\arcdeg$) CME. The projected speed calculated from SECCHI/COR1 Ahead images was 297 km s$^{-1}$ and from SECCHI/COR1 Behind images, it was estimated as 328 km s$^{-1}$. The CME first appeared in SECCHI/COR1 Ahead FOV at 19:25 UT and then in SECCHI/COR1 Behind FOV at 20:05 UT.

In the J-map, tracking of feature could be done up to 35$\arcdeg$ elongation for STEREO A and 30$\arcdeg$ for STEREO B. In the J-map constructed from images taken by STEREO A spacecraft, there are two nearly horizontal and one slanted lines in HI2 FOV which start at elongation angle of 35.2$\arcdeg$, 49.4$\arcdeg$ and 69.6$\arcdeg$ respectively. These lines appeared due to planets Venus, Earth and Jupiter respectively in HI2 A FOV. In the J-map constructed from STEREO B images, two horizontal lines which start at elongation angle of 39.8$\arcdeg$ and 47.9$\arcdeg$ respectively are due to appearance of planets Venus and Earth respectively in HI2 B FOV. Obtained kinematics for this CME is shown in Figure 17, where the gap of estimated parameters occurs due to existence of singularity in triangulation scheme.

We used the drag based model (DBM) developed by \citet{Vrsnak2012} to estimate the arrival time of CME. For this event, we used the DBM with take-off distance as 120.65 R$_\sun$, take-off date and time of CME as 13 October 2010 at 09:33 UT, take-off velocity as 354 km s$^{-1}$, ambient solar wind speed as 350 km s$^{-1}$ and drag parameter 0.2 $\times$ 10.7 km$^{-1}$ - 2 $\times$ 10.7 km$^{-1}$ and obtained its arrival time at 11:40 - 11:45 UT on 15 October and transit velocity 354 - 353 km s$^{-1}$ at L1. By extrapolating the second order polynomial fit for distance, predicted arrival time of CME  at L1 is 14 October at 22:53 UT. From the 3D reconstruction of CME leading edge using tie-pointing method on SECCHI/COR2 data, estimated true velocity was found to be 565.8 km s$^{-1}$ at 14.5 R$_\sun$ at 06:50 UT on 11 October 2010. Assuming that the speed of CME is constant beyond COR2 FOV, the predicted arrival time is at 02:33 UT on 14 October 2010 at L1 ($\approx$ 1 AU).

We analyzed the \textit{in-situ} data and identified the CME boundary which is shown in Figure 18. In this figure, CME sheath arrival is marked by the first vertical dashed line (red) at 04:30 UT on 15 October 2010. Second vertical dashed line (red) marks the trailing edge of CME sheath region at 01:38 UT on 16 October 2010. Hatched line (blue) marks the predicted arrival time (with uncertainties) obtained using DBM.

\subsection{26 October 2010 CME} This CME was observed by both the STEREO spacecraft in COR1 images on 26 October. This CME was also observed in SOHO/LASCO images around 01:36 UT with angular width of 83$\arcdeg$ and had projected linear speed of 214 km s$^{-1}$ at position angle of 210$\arcdeg$.

Independent elongation angle of a tracked feature from two different vantage points is extracted using J-maps. In both J-maps, leading edge of bright feature having positive slope which reveal the CME propagation could be tracked up to 28$\arcdeg$ without ambiguity. For comparison, on 26 October in HI2 A FOV, planets Venus, Earth and Jupiter are observed at elongation angle of 38.3$\arcdeg$, 48.9$\arcdeg$ and 55.8$\arcdeg$ respectively. In HI2 B FOV, planets Venus and Earth are observed at elongation angle of 36.9$\arcdeg$ and 46.7$\arcdeg$ respectively. Figure 19 shows the kinematics of CME of 26 October 2010, which is obtained by implementing GT technique \citep{Liu2010} with the independent elongations estimated from two different viewpoints as inputs. There is a gap in estimated parameters (Figure 19) which is due to existence of singularity in triangulation scheme.

We applied the tie-pointing method for 3D reconstruction of a feature along the identified bright blob behind the leading edge in both set of calibrated and background subtracted images obtained by SECCHI/COR2, to estimate its true kinematics in COR2 FOV. Estimated latitude and longitude show that CME is Earth-directed. The speed of the CME using 3D reconstruction at outer edge of COR2 FOV is about 600 km s$^{-1}$ with acceleration of 30 m s$^{-2}$. Assuming that the speed of CME is constant beyond this distance,  the predicted arrival time of CME at L1 is at 07:45 UT on 29 October 2010. We used the DBM with take-off date and time as 28 October 2010 at 13:57 UT, take-off distance as 110.5 R$_\sun$, take-off velocity as 341 km s$^{-1}$, the ambient solar wind speed as 350 km s$^{-1}$ with drag parameter value 0.2 $\times$ 10$^{-7}$, and the resulting CME arrival time is at 23:45 UT on 30 October 2010 and transit speed is 341 km s$^{-1}$ at L1. Keeping all the input parameters same but with different drag parameter value, 2.0 $\times$ 10$^{-7}$ km$^{-1}$, the CME arrival time is at 23:35 UT on 30 October 2010 and transit speed is 343 km s$^{-1}$ at L1. Predicted arrival time of CME at L1, obtained by extrapolating a second order polynomial fit for distance is at 08:32 UT on 30 October.

We analyzed the \textit{in-situ} observations of solar wind taken nearly at 1 AU and identification of CME is shown in Figure 20.  
In this figure, there is a sudden enhancement of density, temperature and velocity at 10:32 on 30 October which marks the arrival of shock and is indicated by the first dashed vertical line (red) from the left. The hatched line (blue) mark the predicted arrival time (with uncertainty) of CME using DBM. From the left, second and fourth vertical dashed line (red) mark the leading and trailing edge of CME. Leading and trailing edge of magnetic cloud \citep{Klein1982, Lepping1990} is marked by the third and fourth vertical dashed line (red) at 01:30 and 21:35 UT 01 November, respectively. The region of CME has low proton beta \citep{Burlaga1981, Cane2003}, decrease in proton density \citep{Richardson2000}, decrease in proton temperature \citep{Gosling1973}, monotonic decrease in proton velocity \citep{Klein1982} and enhanced alpha to proton ratio \citep{Borrini1982} which are also observed in \textit{in-situ} data and marked with CME in Figure 20.

\section{Results and Discussions} We studied kinematics for 8 selected CMEs by implementing GT technique \citep{Liu2010} which uses elongation (derived from J-maps) as input. We tracked the positively inclined bright feature in J-maps which are considered as  enhanced density regions of a CME moving along the ecliptic. The arrival time of these bright feature is expected to match with the arrival of enhanced density feature in \textit{in-situ} observations. Therefore, in the present study, we define the actual arrival time of a CME as the time when the first density peak is observed in \textit{in-situ} measurements taken at L1 point.

We predicted the arrival time (T$_{arr}$) of CME at L1 point using three different approaches, described  in Section 3.1.3. The arrival time predicted from the two different approaches (DBM \& polynomial fit) which use inputs of derived parameters from GT with J-maps, were compared with the actual arrival time of enhanced density feature at L1 point. For each CME studied here, the actual arrival time and errors in predicted arrival time using these two approaches are shown in Table 1. The predicted arrival time of a CME using DBM is shown in the third column of Table 1, corresponding to the two extreme values of the range of drag parameter. Predicted transit velocity (v$_{1}$) of a CME at L1 is also compared with the actual measured velocity (5th column) \textit{in-situ}, at the time of the arrival of the enhanced density feature. Negative (positive) values of errors in arrival time indicate that predicted arrival is earlier (later) than the actual arrival time. Negative (positive) value of error in transit velocity of CME at L1 indicate that predicted transit velocity of CME at L1 is lesser (greater) than the transit velocity measured \textit{in-situ}.

In order to examine the efficacy of the third approach, one must be certain that the arrival time of tracked feature at L1 is correctly marked . It is generally difficult to mark the actual (reference) arrival time of remotely observed (in COR2) feature at L1 by analyzing the \textit{in-situ} data. This difficulty arises due to uncertainty in one-to-one identification of remotely observed structures with \textit{in-situ} observed structures. In the previous studies, various authors have taken different signatures of CMEs near 1 AU as the reference to correlate the remote sensing observations with their \textit{in-situ} observations. For instance, in the derivation of empirical CME arrival time prediction  model, \citet{Gopalswamy2000} adopted the start time of the magnetic cloud and low proton beta ($<$ 1) as a reference for the actual arrival time of CME. In another study, \citet{Schwenn2005}, interplanetary shock was taken as a reliable \textit{in-situ} signature for a CME. In the study by \citet{Zhang2003}, the minimum Dst index of the associated geomagnetic storm was considered as ICME arrival time. \citet{Kilpua2012} considered the arrival time of CME leading edge at \textit{in-situ} spacecraft as the arrival of CME. In our study, the majority of selected CMEs are slow and do not drive an interplanetary shock. Further, only three events i.e. 12 December 2008, 08 April 2010 and 26 October 2010 could be categorized as magnetic clouds in \textit{in-situ} measurements. For all the CMEs studied in this paper, except 26 October 2010, we have estimated the true speed of a feature along their leading edge in COR2 field-of-view by implementing the tie-pointing method. Also, for all the CMEs, except 10 October 2010, leading and trailing boundary of the CMEs could be identified very well using \textit{in-situ} data.  Therefore, for our study we take the CME leading edge in \textit{in-situ} data as a reference for CME arrival (of a selected feature in COR2) as the most appropriate approach for comparing the predicted arrival time of CME (selected feature in COR2) with the actual arrival time of CME leading edge at L1. The error in predicted arrival time of CME using true speed (estimated in COR2 FOV) is shown in Table 2. In this table, true speed in COR2 FOV and measured speed of CME leading edge at L1 is also shown. In this table, the CME of 10 October 2010 is not included because the CME boundaries could not be identified in \textit{in-situ} data, as it seems probable that only the CME flank was encountered by the spacecraft. It would have been possible to confirm the above, by discussing multi-point observations of this CME, if the STEREO spacecraft had a smaller separation angle at this time, similar to the study carried out by \citet{Kilpua2011}.

From the Table 1 and 2, it is clear that, in general, more accurate prediction of CME arrival time is possible by using DBM combined with estimated kinematics by GT technique. It is also obvious from Table 2 that, for 03 April and 08 April 2010 CMEs, implementation of tie-pointing technique gives better accuracy in arrival time prediction. In these cases, CMEs transit speeds at L1 are approximately equal to measured of CMEs speeds in COR2 FOV, which shows that speeds of CME did not change significantly during its propagation in the heliosphere. Therefore, in these particular cases, observed CME speed in COR2 FOV is sufficient to predict their arrival time near 1 AU with reasonably good accuracy. Our analysis of 03 April CME shows that CME speed is partly governed by high speed stream from a coronal hole located in a geoeffective location on the Sun. The estimated speed ($\approx$ 816 km s$^{-1}$) of 03 April CME in COR2 FOV  and measured speed ($\approx$ 800 km s$^{-1}$) \textit{in-situ} highlights the weak drag force experienced by this CME throughout its journey up to 1 AU. Therefore, CME speed has not tends towards very close to ambient solar wind speed even up to 1 AU. These findings are in good agreement with \citet{Temmer2011}.

Prediction of arrival time by extrapolating the fitted second order polynomial for estimated distance is also better than the prediction made using only the true speed estimated in COR2 FOV. The decelerating trend of 12 December 2008 CME in inner heliosphere is in good agreement with the \citet{Liu2010}. It is also noticed that using extrapolation, the error in predicted arrival time is less if a CME is tracked up to very large distances in heliosphere (using HI), as in the case of 07 February 2010 CME where the J-map allowed tracking of CME features up to nearly 50$\arcdeg$ elongation ($\approx$ 170 R$_\sun$ in this case). High speed ($\approx$ 867 km s$^{-1}$) CME of 12 February 2010 shows a significant deceleration continuously in the heliosphere. This is in agreement with the results of previous studies which demonstrate that drag force plays an important role in shaping the CME dynamics \citep{Lindsay1999, Gopalswamy2001, Manoharan2006, Vrsnak2007}.

In our study, CMEs could be tracked in HI images up to a large elongation angle ($\approx$ 35$\arcdeg$) however relating the tracked feature to \textit{in-situ} features observed is often challenging. In GT technique, although, we assume that same enhanced density structure is tracked using J-maps from both STEREO spacecraft and also in each consecutive image but these assumptions are not truly valid for observed CMEs. Even if these assumptions are considered valid, it is possible that single \textit{in-situ} spacecraft is unable to sample the tracked feature. In this worse situation, relating remotely observed tracked feature with \textit{in-situ} observations will lead to incorrect interpretation. Since, the location of \textit{in-situ} spacecraft with respect to CME structure, decides which part of CME will be intercepted by the \textit{in-situ} spacecraft, it may be more appropriate to take the geometry of CMEs also into account in GT scheme. In the present work, we made the J-map along the ecliptic plane as the estimated velocity in this plane is more suitable than radial velocity at other position angles for estimating the arrival time of CMEs at L1 point. But, even in the ecliptic plane, J-map gives information about only a part of CME structure which has different velocity at different longitudes. The time taken by different parts of a CME to traverse a fixed radial distance from the Sun is different and it is minimum for the apex of a CME \citep{Schwenn2005}. If the tracked feature happens to be the apex of a CME which is moving at different direction (longitude) from Sun-L1 line, then the predicted arrival time of this CME using the estimated speed of tracked feature will be earlier than the actual arrival time of CME at L1 point.  Therefore, to obtain the actual arrival time and transit velocity of a CME passing through the in-situ spacecraft, it is required to take into account the propagation direction of tracked feature. \citet{Mostl2011} showed that the speed of a CME flank measured at a given angle $\theta$ to the CME apex is reduced by a factor of cos($\theta$) for a circular geometry of CME. Therefore, it seems reasonable that transit speed measured \textit{in-situ} and arrival time of CME at L1 should be compared with the corresponding corrected speed of tracked feature along the Sun-L1 line and the arrival time respectively. However, it is to be noted that if the apex of CME moves with a linear speed of 400 km s$^{-1}$ at 10$\arcdeg$ to the Sun-L1 line, the correction in actual arrival time of CME at L1 is only approximately 2 hours for a fixed distance of nearly 1 AU. It sounds appropriate to take the geometry of CMEs also in triangulation scheme. However, idealistic assumptions on geometry also are far from the real structure. Further, CME shapes can be distorted in heliosphere by solar wind, IP shocks and CMEs interactions. Therefore, one needs to cautions as the assumptions made for geometry of CMEs may result in new sources of errors.

In spite of various assumptions made in the DBM \citep{Vrsnak2012} and in the GT technique \citep{Liu2010}, prediction of arrival time of CMEs is found within acceptable error. It must be mentioned here, that the estimated velocity profiles of the selected CMEs, show small apparent acceleration and deceleration for few hours which do not seem to be real. We believe that these may be due to errors in manual tracking of a CME feature using J-maps and extracting the elongation angles. In the kinematics plot, we have shown the errors bars with vertical red lines. However, these do not denote the actual errors in triangulation scheme, but is representative of sensitivity of technique to elongation uncertainties \citep{Liu2010}. Considering that elongation angles determined for each pixel in level 0 data for COR2 and level 2 data for HI1 and HI2 are quite accurate, then the minimum uncertainty in elongation will be determined by the resolution in elongation which is used to construct the J-maps. However, as mentioned above, the actual error in this technique owes to the manual error in tracking the bright points using J-maps. The assessment of this error is possible by repeating the manual tracking of feature several times and comparing the derived parameters. However, one should also ensure that same feature is tracked continuously in the J-maps. This is often difficult due to different sensitivities of COR2, HI1 and HI2 imaging instruments covering a wide range of elongation angle. The actual sources of error in GT technique is inherent in the assumption that CME feature is a single point and the same point is being tracked continuously from both twin STEREO spacecraft. If the assumption breaks (likely $\sim$ at larger elongations) at some segment of journey in the interplanetary medium, the CME kinematics will not be correctly estimated.

It may also be noted that the density distribution along the line of sight is not well known, as we project 3D structure of CME features on to a 2D image.  Relating remote sensing observations to \textit{in-situ} measurements of CMEs is often uncertain, because bright feature observed in J-maps is due to the contribution of intensity along the entire depth of line-of-sight but ACE or WIND density measurements near 1 AU are only along the Sun-Earth line. In spite of these constraints on observations and assumptions in implemented techniques, our  study shows a fair agreement in relating remotely observed features with \textit{in-situ} measurements, by tracking the CMEs up to large distances in the heliosphere. The efficacy of any forecasting scheme for CME arrival time must be validated with real time data so that the results are unbiased. Further, we aim to implement other techniques; e.g. Fixed $\phi$: \citet{Kahler2007}, Harmonic Mean Fitting: \citet{Lugaz2009}, Triangulation with Harmonic Mean: \citet{Lugaz2010}, SSE Method: \citet{Davies2012}; on these CMEs to ascertain and compare the kinematics obtained in the heliosphere.

\section{Summary}
We have studied kinematics of eight CMEs by exploiting the STEREO COR2 and HI observations. Speeds of selected CMEs in our study range from low ($\approx$ {335 km s$^{-1}$}) to high ($\approx$ {870 km s$^{-1}$}) in the coronagraphic (COR2) field of view. We obtained a good agreement (within $\approx{100}$ km s$^{-1}$) between speed calculated using tie-pointing method of 3D reconstruction and speed derived by implementing GT technique using J-maps in the COR2 FOV. The difference in the speeds using two different techniques can occur due to estimation of speeds of different features at different latitudes.

Based on our prediction of arrival time of CME at L1 using three different approaches, our study reveals that use of GT technique on HI data combined with DBM gives a better prediction of CME arrival time, with an error of 3 to 9 hours and transit velocity near 1 AU with an error ranging between 25 to 120 km s$^{-1}$. Our study also shows that using true speed of CMEs determined at the farthest point in COR2 FOV and assuming that the speed remains constant for the remaining distance, i.e. up to 1 AU, may not be sufficient to predict their arrival time accurately for a majority of events. Also, it is worth to mention that a small change in estimated speed in COR2 FOV may result in a large variation in predicted arrival time at near 1 AU, which may be due to large distance between COR2 FOV and near 1 AU. For a comparison, change in estimated speed from 400 to 450 km s$^{-1}$ can give a  difference in estimated arrival time of 10.5 hours near 1 AU. However, for fast CMEs, this variation in arrival time using slightly different estimated true speeds, will be minimized. Our study shows that exploiting wide angle imaging data from the Heliospheric Imager (HI) with GT technique combined with DBM, helps to understand the relationship between remote sensing and \textit{in-situ} observations, CME acceleration and deceleration beyond COR2 FOV and role of ambient solar wind in depth. Our technique also reveals the non-radial motion of CMEs even far from the Sun from the estimation of propagation direction.

\subsection *{Acknowledgments}
\acknowledgments
We are thankful to J. A. Davies, Anand D. Joshi and Y. Liu for their support with programming in the preparation of J-maps and helpful discussions on singularities in geometric triangulation technique. We thank the STEREO/SECCHI, ACE \& WIND team for their open data policy. The work by NS partially contributes to the research for European Union Seventh Framework Programme (FP7/2007-2013) for the Coronal Mass Ejections and Solar Energetic Particles (COMESEP) project under Grant Agreement No. 263252. The authors gratefully acknowledge the U K Solar System Data Centre for providing the calibrated and background subtracted STEREO/HI data.

\clearpage

\bibliographystyle{apj}
\bibliography{wag_paper}

\begin{thebibliography}{84}
\expandafter\ifx\csname natexlab\endcsname\relax\def\natexlab#1{#1}\fi

\bibitem[{{Andrews} {et~al.}(1999){Andrews}, {Wang}, \& {Wu}}]{Andrews1999}
{Andrews}, M.~D., {Wang}, A.-H., \& {Wu}, S.~T. 1999, \solphys, 187, 427

\bibitem[{{Billings}(1966)}]{Billings1966}
{Billings}, D.~E. 1966, {A guide to the solar corona} (Academic Press, San
  Diego)

\bibitem[{{Borrini} {et~al.}(1982){Borrini}, {Gosling}, {Bame}, \&
  {Feldman}}]{Borrini1982}
{Borrini}, G., {Gosling}, J.~T., {Bame}, S.~J., \& {Feldman}, W.~C. 1982, \jgr,
  87, 7370

\bibitem[{{Boursier} {et~al.}(2009){Boursier}, {Lamy}, \&
  {Llebaria}}]{Boursier2009}
{Boursier}, Y., {Lamy}, P., \& {Llebaria}, A. 2009, \solphys, 256, 131

\bibitem[{{Brueckner} {et~al.}(1995){Brueckner}, {Howard}, {Koomen},
  {Korendyke}, {Michels}, {Moses}, {Socker}, {Dere}, {Lamy}, {Llebaria},
  {Bout}, {Schwenn}, {Simnett}, {Bedford}, \& {Eyles}}]{Brueckner1995}
{Brueckner}, G.~E., {Howard}, R.~A., {Koomen}, M.~J., {et~al.} 1995, \solphys,
  162, 357

\bibitem[{{Burlaga} {et~al.}(1981){Burlaga}, {Sittler}, {Mariani}, \&
  {Schwenn}}]{Burlaga1981}
{Burlaga}, L., {Sittler}, E., {Mariani}, F., \& {Schwenn}, R. 1981, \jgr, 86,
  6673

\bibitem[{{Byrne} {et~al.}(2010){Byrne}, {Maloney}, {McAteer}, {Refojo}, \&
  {Gallagher}}]{Byrne2010}
{Byrne}, J.~P., {Maloney}, S.~A., {McAteer}, R.~T.~J., {Refojo}, J.~M., \&
  {Gallagher}, P.~T. 2010, Nature Communications, 1

\bibitem[{{Cane} \& {Richardson}(2003)}]{Cane2003}
{Cane}, H.~V., \& {Richardson}, I.~G. 2003, Journal of Geophysical Research
  (Space Physics), 108, 1156

\bibitem[{{Cargill}(2004)}]{Cargill2004}
{Cargill}, P.~J. 2004, \solphys, 221, 135

\bibitem[{{Colaninno} \& {Vourlidas}(2009)}]{Colaninno2009}
{Colaninno}, R.~C., \& {Vourlidas}, A. 2009, \apj, 698, 852

\bibitem[{{Davies} {et~al.}(2009){Davies}, {Harrison}, {Rouillard}, {Sheeley},
  {Perry}, {Bewsher}, {Davis}, {Eyles}, {Crothers}, \& {Brown}}]{Davies2009}
{Davies}, J.~A., {Harrison}, R.~A., {Rouillard}, A.~P., {et~al.} 2009, \grl,
  36, 2102

\bibitem[{{Davies} {et~al.}(2012){Davies}, {Harrison}, {Perry}, {M{\"o}stl},
  {Lugaz}, {Rollett}, {Davis}, {Crothers}, {Temmer}, {Eyles}, \&
  {Savani}}]{Davies2012}
{Davies}, J.~A., {Harrison}, R.~A., {Perry}, C.~H., {et~al.} 2012, \apj, 750,
  23

\bibitem[{{Davis} {et~al.}(2009){Davis}, {Davies}, {Lockwood}, {Rouillard},
  {Eyles}, \& {Harrison}}]{Davis2009}
{Davis}, C.~J., {Davies}, J.~A., {Lockwood}, M., {et~al.} 2009, \grl, 36, 8102

\bibitem[{{Dryer}(1994)}]{Dryer1994}
{Dryer}, M. 1994, \ssr, 67, 363

\bibitem[{{Dryer} {et~al.}(2004){Dryer}, {Smith}, {Fry}, {Sun}, {Deehr}, \&
  {Akasofu}}]{Dryer2004}
{Dryer}, M., {Smith}, Z., {Fry}, C.~D., {et~al.} 2004, Space Weather, 2, 9001

\bibitem[{{Dungey}(1961)}]{Dungey1961}
{Dungey}, J.~W. 1961, Physical Review Letters, 6, 47

\bibitem[{{Echer} {et~al.}(2008){Echer}, {Gonzalez}, {Tsurutani}, \&
  {Gonzalez}}]{Echer2008}
{Echer}, E., {Gonzalez}, W.~D., {Tsurutani}, B.~T., \& {Gonzalez}, A.~L.~C.
  2008, Journal of Geophysical Research (Space Physics), 113, 5221

\bibitem[{{Eyles} {et~al.}(2009){Eyles}, {Harrison}, {Davis}, {Waltham},
  {Shaughnessy}, {Mapson-Menard}, {Bewsher}, {Crothers}, {Davies}, {Simnett},
  {Howard}, {Moses}, {Newmark}, {Socker}, {Halain}, {Defise}, {Mazy}, \&
  {Rochus}}]{Eyles2009}
{Eyles}, C.~J., {Harrison}, R.~A., {Davis}, C.~J., {et~al.} 2009, \solphys,
  254, 387

\bibitem[{{Feng} {et~al.}(2009){Feng}, {Zhang}, {Sun}, {Dryer}, {Fry}, \&
  {Deehr}}]{Feng2009}
{Feng}, X.~S., {Zhang}, Y., {Sun}, W., {et~al.} 2009, Journal of Geophysical
  Research (Space Physics), 114, 1101

\bibitem[{{Gopalswamy} {et~al.}(2000){Gopalswamy}, {Lara}, {Lepping}, {Kaiser},
  {Berdichevsky}, \& {St.~Cyr}}]{Gopalswamy2000}
{Gopalswamy}, N., {Lara}, A., {Lepping}, R.~P., {et~al.} 2000, \grl, 27, 145

\bibitem[{{Gopalswamy} {et~al.}(2005){Gopalswamy}, {Lara}, {Manoharan}, \&
  {Howard}}]{Gopalswamy2005}
{Gopalswamy}, N., {Lara}, A., {Manoharan}, P.~K., \& {Howard}, R.~A. 2005,
  Advances in Space Research, 36, 2289

\bibitem[{{Gopalswamy} {et~al.}(2001){Gopalswamy}, {Yashiro}, {Kaiser},
  {Howard}, \& {Bougeret}}]{Gopalswamy2001}
{Gopalswamy}, N., {Yashiro}, S., {Kaiser}, M.~L., {Howard}, R.~A., \&
  {Bougeret}, J.-L. 2001, \jgr, 106, 29219

\bibitem[{{Gosling} {et~al.}(1990){Gosling}, {Bame}, {McComas}, \&
  {Phillips}}]{Gosling1990}
{Gosling}, J.~T., {Bame}, S.~J., {McComas}, D.~J., \& {Phillips}, J.~L. 1990,
  \grl, 17, 901

\bibitem[{{Gosling} {et~al.}(1991){Gosling}, {McComas}, {Phillips}, \&
  {Bame}}]{Gosling1991}
{Gosling}, J.~T., {McComas}, D.~J., {Phillips}, J.~L., \& {Bame}, S.~J. 1991,
  \jgr, 96, 7831

\bibitem[{{Gosling} {et~al.}(1973){Gosling}, {Pizzo}, \& {Bame}}]{Gosling1973}
{Gosling}, J.~T., {Pizzo}, V., \& {Bame}, S.~J. 1973, \jgr, 78, 2001

\bibitem[{{Harrison} {et~al.}(2012){Harrison}, {Davies}, {M{\"o}stl}, {Liu},
  {Temmer}, {Bisi}, {Eastwood}, {de Koning}, {Nitta}, {Rollett}, {Farrugia},
  {Forsyth}, {Jackson}, {Jensen}, {Kilpua}, {Odstrcil}, \&
  {Webb}}]{Harrison2012}
{Harrison}, R.~A., {Davies}, J.~A., {M{\"o}stl}, C., {et~al.} 2012, \apj, 750,
  45

\bibitem[{{Howard} {et~al.}(2008){Howard}, {Moses}, {Vourlidas}, {Newmark},
  {Socker}, {Plunkett}, {Korendyke}, {Cook}, {Hurley}, {Davila}, {Thompson},
  {St Cyr}, {Mentzell}, {Mehalick}, {Lemen}, {Wuelser}, {Duncan}, {Tarbell},
  {Wolfson}, {Moore}, {Harrison}, {Waltham}, {Lang}, {Davis}, {Eyles},
  {Mapson-Menard}, {Simnett}, {Halain}, {Defise}, {Mazy}, {Rochus}, {Mercier},
  {Ravet}, {Delmotte}, {Auchere}, {Delaboudiniere}, {Bothmer}, {Deutsch},
  {Wang}, {Rich}, {Cooper}, {Stephens}, {Maahs}, {Baugh}, {McMullin}, \&
  {Carter}}]{Howard2008}
{Howard}, R.~A., {Moses}, J.~D., {Vourlidas}, A., {et~al.} 2008, \ssr, 136, 67

\bibitem[{{Howard} \& {Tappin}(2009{\natexlab{a}})}]{Howard2009}
{Howard}, T.~A., \& {Tappin}, S.~J. 2009{\natexlab{a}}, \ssr, 147, 31

\bibitem[{{Howard} \& {Tappin}(2009{\natexlab{b}})}]{Howard2009a}
---. 2009{\natexlab{b}}, \ssr, 147, 89

\bibitem[{{Hundhausen} {et~al.}(1984){Hundhausen}, {Sawyer}, {House}, {Illing},
  \& {Wagner}}]{Hundhausen1984}
{Hundhausen}, A.~J., {Sawyer}, C.~B., {House}, L., {Illing}, R.~M.~E., \&
  {Wagner}, W.~J. 1984, \jgr, 89, 2639

\bibitem[{{Inhester}(2006)}]{Inhester2006}
{Inhester}, B. 2006, ArXiv Astrophysics e-prints

\bibitem[{{Kahler} \& {Webb}(2007)}]{Kahler2007}
{Kahler}, S.~W., \& {Webb}, D.~F. 2007, Journal of Geophysical Research (Space
  Physics), 112, 9103

\bibitem[{{Kaiser} {et~al.}(2008){Kaiser}, {Kucera}, {Davila}, {St.~Cyr},
  {Guhathakurta}, \& {Christian}}]{Kaiser2008}
{Kaiser}, M.~L., {Kucera}, T.~A., {Davila}, J.~M., {et~al.} 2008, \ssr, 136, 5

\bibitem[{{Kilpua} {et~al.}(2011){Kilpua}, {Jian}, {Li}, {Luhmann}, \&
  {Russell}}]{Kilpua2011}
{Kilpua}, E.~K.~J., {Jian}, L.~K., {Li}, Y., {Luhmann}, J.~G., \& {Russell},
  C.~T. 2011, Journal of Atmospheric and Solar-Terrestrial Physics, 73, 1228

\bibitem[{{Kilpua} {et~al.}(2012){Kilpua}, {Mierla}, {Rodriguez}, {Zhukov},
  {Srivastava}, \& {West}}]{Kilpua2012}
{Kilpua}, E.~K.~J., {Mierla}, M., {Rodriguez}, L., {et~al.} 2012, \solphys,
  279, 477

\bibitem[{{Klein} \& {Burlaga}(1982)}]{Klein1982}
{Klein}, L.~W., \& {Burlaga}, L.~F. 1982, \jgr, 87, 613

\bibitem[{{Lara} \& {Borgazzi}(2009)}]{Lara2009}
{Lara}, A., \& {Borgazzi}, A.~I. 2009, in IAU Symposium, Vol. 257, IAU
  Symposium, ed. N.~{Gopalswamy} \& D.~F. {Webb}, 287--290

\bibitem[{{Leblanc} {et~al.}(1998){Leblanc}, {Dulk}, \&
  {Bougeret}}]{Leblanc1998}
{Leblanc}, Y., {Dulk}, G.~A., \& {Bougeret}, J.-L. 1998, \solphys, 183, 165

\bibitem[{{Lepping} {et~al.}(1990){Lepping}, {Burlaga}, \&
  {Jones}}]{Lepping1990}
{Lepping}, R.~P., {Burlaga}, L.~F., \& {Jones}, J.~A. 1990, \jgr, 95, 11957

\bibitem[{{Lindsay} {et~al.}(1999){Lindsay}, {Luhmann}, {Russell}, \&
  {Gosling}}]{Lindsay1999}
{Lindsay}, G.~M., {Luhmann}, J.~G., {Russell}, C.~T., \& {Gosling}, J.~T. 1999,
  \jgr, 104, 12515

\bibitem[{{Liu} {et~al.}(2010){Liu}, {Davies}, {Luhmann}, {Vourlidas}, {Bale},
  \& {Lin}}]{Liu2010}
{Liu}, Y., {Davies}, J.~A., {Luhmann}, J.~G., {et~al.} 2010, \apjl, 710, L82

\bibitem[{{Liu} {et~al.}(2011){Liu}, {Luhmann}, {Bale}, \& {Lin}}]{Liu2011}
{Liu}, Y., {Luhmann}, J.~G., {Bale}, S.~D., \& {Lin}, R.~P. 2011, \apj, 734, 84

\bibitem[{{Liu} {et~al.}(2012){Liu}, {Luhmann}, {M{\"o}stl},
  {Martinez-Oliveros}, {Bale}, {Lin}, {Harrison}, {Temmer}, {Webb}, \&
  {Odstrcil}}]{Liu2012}
{Liu}, Y.~D., {Luhmann}, J.~G., {M{\"o}stl}, C., {et~al.} 2012, \apjl, 746, L15

\bibitem[{{Lopez}(1987)}]{Lopez1987}
{Lopez}, R.~E. 1987, \jgr, 92, 11189

\bibitem[{{Lopez} \& {Freeman}(1986)}]{Lopez1986}
{Lopez}, R.~E., \& {Freeman}, J.~W. 1986, \jgr, 91, 1701

\bibitem[{{Lugaz} {et~al.}(2012){Lugaz}, {Farrugia}, {Davies}, {M{\"o}stl},
  {Davis}, {Roussev}, \& {Temmer}}]{Lugaz2012}
{Lugaz}, N., {Farrugia}, C.~J., {Davies}, J.~A., {et~al.} 2012, \apj, 759, 68

\bibitem[{{Lugaz} {et~al.}(2010){Lugaz}, {Hernandez-Charpak}, {Roussev},
  {Davis}, {Vourlidas}, \& {Davies}}]{Lugaz2010}
{Lugaz}, N., {Hernandez-Charpak}, J.~N., {Roussev}, I.~I., {et~al.} 2010, \apj,
  715, 493

\bibitem[{{Lugaz} {et~al.}(2009){Lugaz}, {Vourlidas}, \& {Roussev}}]{Lugaz2009}
{Lugaz}, N., {Vourlidas}, A., \& {Roussev}, I.~I. 2009, Annales Geophysicae,
  27, 3479

\bibitem[{{Maloney} \& {Gallagher}(2010)}]{Maloney2010}
{Maloney}, S.~A., \& {Gallagher}, P.~T. 2010, \apjl, 724, L127

\bibitem[{{Manchester} {et~al.}(2004){Manchester}, {Gombosi}, {Roussev},
  {Ridley}, {De Zeeuw}, {Sokolov}, {Powell}, \& {T{\'o}th}}]{Manchester2004}
{Manchester}, W.~B., {Gombosi}, T.~I., {Roussev}, I., {et~al.} 2004, Journal of
  Geophysical Research (Space Physics), 109, 2107

\bibitem[{{Manoharan}(2006)}]{Manoharan2006}
{Manoharan}, P.~K. 2006, \solphys, 235, 345

\bibitem[{{Mierla} {et~al.}(2009){Mierla}, {Inhester}, {Marqu{\'e}},
  {Rodriguez}, {Gissot}, {Zhukov}, {Berghmans}, \& {Davila}}]{Mierla2009}
{Mierla}, M., {Inhester}, B., {Marqu{\'e}}, C., {et~al.} 2009, \solphys, 259,
  123

\bibitem[{{Mierla} {et~al.}(2008){Mierla}, {Davila}, {Thompson}, {Inhester},
  {Srivastava}, {Kramar}, {St.~Cyr}, {Stenborg}, \& {Howard}}]{Mierla2008}
{Mierla}, M., {Davila}, J., {Thompson}, W., {et~al.} 2008, \solphys, 252, 385

\bibitem[{{Moran} \& {Davila}(2004)}]{Moran2004}
{Moran}, T.~G., \& {Davila}, J.~M. 2004, Science, 305, 66

\bibitem[{{M{\"o}stl} {et~al.}(2011){M{\"o}stl}, {Rollett}, {Lugaz},
  {Farrugia}, {Davies}, {Temmer}, {Veronig}, {Harrison}, {Crothers}, {Luhmann},
  {Galvin}, {Zhang}, {Baumjohann}, \& {Biernat}}]{Mostl2011}
{M{\"o}stl}, C., {Rollett}, T., {Lugaz}, N., {et~al.} 2011, \apj, 741, 34

\bibitem[{{Odstrcil} {et~al.}(2004){Odstrcil}, {Riley}, \&
  {Zhao}}]{Odstrcil2004}
{Odstrcil}, D., {Riley}, P., \& {Zhao}, X.~P. 2004, Journal of Geophysical
  Research (Space Physics), 109, 2116

\bibitem[{{Ogilvie} {et~al.}(1995){Ogilvie}, {Chornay}, {Fritzenreiter},
  {Hunsaker}, {Keller}, {Lobell}, {Miller}, {Scudder}, {Sittler}, {Torbert},
  {Bodet}, {Needell}, {Lazarus}, {Steinberg}, {Tappan}, {Mavretic}, \&
  {Gergin}}]{Ogilvie1995}
{Ogilvie}, K.~W., {Chornay}, D.~J., {Fritzenreiter}, R.~J., {et~al.} 1995,
  \ssr, 71, 55

\bibitem[{{Richardson} {et~al.}(2000){Richardson}, {Berdichevsky}, {Desch}, \&
  {Farrugia}}]{Richardson2000}
{Richardson}, I.~G., {Berdichevsky}, D., {Desch}, M.~D., \& {Farrugia}, C.~J.
  2000, \grl, 27, 3761

\bibitem[{{Richardson} \& {Cane}(2010)}]{Richardson2010}
{Richardson}, I.~G., \& {Cane}, H.~V. 2010, \solphys, 264, 189

\bibitem[{{Richardson} {et~al.}(2001){Richardson}, {Cliver}, \&
  {Cane}}]{Richardson2001}
{Richardson}, I.~G., {Cliver}, E.~W., \& {Cane}, H.~V. 2001, \grl, 28, 2569

\bibitem[{{Rouillard} {et~al.}(2009){Rouillard}, {Savani}, {Davies}, {Lavraud},
  {Forsyth}, {Morley}, {Opitz}, {Sheeley}, {Burlaga}, {Sauvaud}, {Simunac},
  {Luhmann}, {Galvin}, {Crothers}, {Davis}, {Harrison}, {Lockwood}, {Eyles},
  {Bewsher}, \& {Brown}}]{Rouillard2009}
{Rouillard}, A.~P., {Savani}, N.~P., {Davies}, J.~A., {et~al.} 2009, \solphys,
  256, 307

\bibitem[{{Schwenn} {et~al.}(2005){Schwenn}, {dal Lago}, {Huttunen}, \&
  {Gonzalez}}]{Schwenn2005}
{Schwenn}, R., {dal Lago}, A., {Huttunen}, E., \& {Gonzalez}, W.~D. 2005,
  Annales Geophysicae, 23, 1033

\bibitem[{{Sheeley} {et~al.}(1999){Sheeley}, {Walters}, {Wang}, \&
  {Howard}}]{Sheeley1999}
{Sheeley}, N.~R., {Walters}, J.~H., {Wang}, Y.-M., \& {Howard}, R.~A. 1999,
  \jgr, 104, 24739

\bibitem[{{Sheeley} {et~al.}(1997){Sheeley}, {Wang}, {Hawley}, {Brueckner},
  {Dere}, {Howard}, {Koomen}, {Korendyke}, {Michels}, {Paswaters}, {Socker},
  {St.~Cyr}, {Wang}, {Lamy}, {Llebaria}, {Schwenn}, {Simnett}, {Plunkett}, \&
  {Biesecker}}]{Sheeley1997}
{Sheeley}, Jr., N.~R., {Wang}, Y.-M., {Hawley}, S.~H., {et~al.} 1997, \apj,
  484, 472

\bibitem[{{Smith} {et~al.}(2009){Smith}, {Dryer}, {McKenna-Lawlor}, {Fry},
  {Deehr}, \& {Sun}}]{Smith2009}
{Smith}, Z.~K., {Dryer}, M., {McKenna-Lawlor}, S.~M.~P., {et~al.} 2009, Journal
  of Geophysical Research (Space Physics), 114, 5106

\bibitem[{{Stone} {et~al.}(1998){Stone}, {Frandsen}, {Mewaldt}, {Christian},
  {Margolies}, {Ormes}, \& {Snow}}]{stone1998}
{Stone}, E.~C., {Frandsen}, A.~M., {Mewaldt}, R.~A., {et~al.} 1998, \ssr, 86, 1

\bibitem[{{Tappin}(1987)}]{Tappin1987}
{Tappin}, S.~J. 1987, \planss, 35, 271

\bibitem[{{Temmer} {et~al.}(2011){Temmer}, {Rollett}, {M{\"o}stl}, {Veronig},
  {Vr{\v s}nak}, \& {Odstr{\v c}il}}]{Temmer2011}
{Temmer}, M., {Rollett}, T., {M{\"o}stl}, C., {et~al.} 2011, \apj, 743, 101

\bibitem[{{Thernisien} {et~al.}(2009){Thernisien}, {Vourlidas}, \&
  {Howard}}]{Thernisien2009}
{Thernisien}, A., {Vourlidas}, A., \& {Howard}, R.~A. 2009, \solphys, 256, 111

\bibitem[{{Thompson}(2009)}]{Thompson2009}
{Thompson}, W.~T. 2009, \icarus, 200, 351

\bibitem[{{Vourlidas} \& {Howard}(2006)}]{Vourlidas2006}
{Vourlidas}, A., \& {Howard}, R.~A. 2006, \apj, 642, 1216

\bibitem[{{Vourlidas} {et~al.}(2000){Vourlidas}, {Subramanian}, {Dere}, \&
  {Howard}}]{Vourlidas2000}
{Vourlidas}, A., {Subramanian}, P., {Dere}, K.~P., \& {Howard}, R.~A. 2000,
  \apj, 534, 456

\bibitem[{{Vr{\v s}nak} \& {Gopalswamy}(2002)}]{Vrsnak2002}
{Vr{\v s}nak}, B., \& {Gopalswamy}, N. 2002, Journal of Geophysical Research
  (Space Physics), 107, 1019

\bibitem[{{Vr{\v s}nak} \& {{\v Z}ic}(2007)}]{Vrsnak2007}
{Vr{\v s}nak}, B., \& {{\v Z}ic}, T. 2007, \aap, 472, 937

\bibitem[{{Vr{\v s}nak} {et~al.}(2010){Vr{\v s}nak}, {{\v Z}ic}, {Falkenberg},
  {M{\"o}stl}, {Vennerstrom}, \& {Vrbanec}}]{Vrsnak2010}
{Vr{\v s}nak}, B., {{\v Z}ic}, T., {Falkenberg}, T.~V., {et~al.} 2010, \aap,
  512, A43

\bibitem[{{Vr{\v s}nak} {et~al.}(2012){Vr{\v s}nak}, {{\v Z}ic}, {Vrbanec},
  {Temmer}, {Rollett}, {M{\"o}stl}, {Veronig}, {{\v C}alogovi{\'c}},
  {Dumbovi{\'c}}, {Luli{\'c}}, {Moon}, \& {Shanmugaraju}}]{Vrsnak2012}
{Vr{\v s}nak}, B., {{\v Z}ic}, T., {Vrbanec}, D., {et~al.} 2012, \solphys, 124

\bibitem[{{Webb} \& {Howard}(2012)}]{Webb2012}
{Webb}, D.~F., \& {Howard}, T.~A. 2012, Living Reviews in Solar Physics, 9, 3

\bibitem[{{Wood} {et~al.}(1999){Wood}, {Karovska}, {Chen}, {Brueckner}, {Cook},
  \& {Howard}}]{Wood1999}
{Wood}, B.~E., {Karovska}, M., {Chen}, J., {et~al.} 1999, \apj, 512, 484

\bibitem[{{Xie} {et~al.}(2004){Xie}, {Ofman}, \& {Lawrence}}]{Xie2004}
{Xie}, H., {Ofman}, L., \& {Lawrence}, G. 2004, Journal of Geophysical Research
  (Space Physics), 109, 3109

\bibitem[{{Yashiro} {et~al.}(2004){Yashiro}, {Gopalswamy}, {Michalek},
  {St.~Cyr}, {Plunkett}, {Rich}, \& {Howard}}]{Yashiro2004}
{Yashiro}, S., {Gopalswamy}, N., {Michalek}, G., {et~al.} 2004, Journal of
  Geophysical Research (Space Physics), 109, 7105

\bibitem[{{Zhang} {et~al.}(2003){Zhang}, {Dere}, {Howard}, \&
  {Bothmer}}]{Zhang2003}
{Zhang}, J., {Dere}, K.~P., {Howard}, R.~A., \& {Bothmer}, V. 2003, \apj, 582,
  520

\bibitem[{{Zhang} {et~al.}(2007){Zhang}, {Richardson}, {Webb}, {Gopalswamy},
  {Huttunen}, {Kasper}, {Nitta}, {Poomvises}, {Thompson}, {Wu}, {Yashiro}, \&
  {Zhukov}}]{Zhang2007}
{Zhang}, J., {Richardson}, I.~G., {Webb}, D.~F., {et~al.} 2007, Journal of
  Geophysical Research (Space Physics), 112, 10102

\bibitem[{{Zhao}(1992)}]{Zhao1992}
{Zhao}, X. 1992, \jgr, 97, 15051

\bibitem[{{Zurbuchen} \& {Richardson}(2006)}]{Zurbuchen2006}
{Zurbuchen}, T.~H., \& {Richardson}, I.~G. 2006, \ssr, 123, 31

\end{thebibliography}

\clearpage

\begin{table}
  \centering

{\scriptsize
 \begin{tabular}{|p{1cm}| p{1.4cm}| p{2.3cm}|p{1.5cm}|p{1.3cm}|p{2.0cm}|}
    \hline
 &  & \multicolumn{2}{ c| }{Error in predicted T$_{arr}$ at L1 (hours) } &  &  \\ \cline{3-4} 
CME dates & Actual T$_{arr}$ (Peak density time) (UT) & Kinematics + Drag Based Model [$\gamma$ = 0.2 - 2.0 (10$^{-7}$ km$^{-1}$)]  & Distance + Polynomial fit & Actual v$_{1}$ at L1 (km s$^{-1}$) & Error in predicted v$_{1}$ at L1 (km s$^{-1}$) [$\gamma$ = 0.2 - 2.0 (10$^{-7}$ km$^{-1}$)]  \\ \hline 

12 Dec 2008 & 16 Dec 23:50 & -3.4 to -3.9 & +6.5  & 356  & -25 to -18  \\ \hline
07 Feb 2010 & 11 Feb 02:05 & -4.3 to -3.2 & -1.2  & 370  & +72 to +23  \\ \hline
12 Feb 2010 & 15 Feb 23:15 & -8.7 to -7.9 & -7.1  & 320  & +122 to +81 \\ \hline
14 Mar 2010 & 17 Mar 21:45 & -0.6 to +3.2 & -5.4  & 453  & -16 to -75  \\ \hline
03 Apr 2010 & 05 Apr 12:00 & +5.5       & -3.0    & 720  & -96         \\ \hline
08 Apr 2010 & 11 Apr 14:10 & -4.4 to -1.2 & -7.6  & 426  & +85 to -24  \\ \hline
10 Oct 2010 & 15 Oct 06:05 & +5.5 to +5.6 & -7.2  & 300  & +54 to +53  \\ \hline
26 Oct 2010 & 31 Oct 03:30 & -3.7 to -4.0 & -18.9 & 365  & -24 to -22  \\ \hline 

 \end{tabular}
}
\caption{The errors in predicted arrival time using two approaches are shown in column 3rd \& 4th and errors in predicted transit velocity (column 6th). Actual arrival time and transit velocity of tracked feature at L1 point is shown in column 2nd and 5th, respectively.}
\end{table}

\begin{table}
  \centering

{\scriptsize
 \begin{tabular}{|p{2cm}| p{2cm}| p{1.5cm}|p{1.7cm}|p{1.2cm}|}
    \hline
CME dates &  Actual arrival time (UT) of CME leading edge at L1 &  Error in predicted arrival time & Measured velocity of CME leading edge at L1 & Velocity (km s$^{-1}$) in COR2 FOV  \\ \hline

12 Dec 2008 & 17 Dec 04:39 & -26.9 & 365 & 453 \\ \hline
07 Feb 2010 & 11 Feb 12:47 & -21.8 & 360 & 480 \\ \hline
12 Feb 2010 & 16 Feb 04:32 & -41.5 & 310 & 867  \\ \hline
14 Mar 2010 & 17 Mar 21:19 & +27 & 450 & 335  \\ \hline
03 Apr 2010 & 05 Apr 13:43 & -2.3 & 800 & 816 \\ \hline
08 Apr 2010 & 12 Apr 02:10 & -9.5 & 410 & 478  \\ \hline
26 Oct 2010 & 31 Oct 06:30 & -46.7 & 365 & 600 \\ \hline 

 \end{tabular}
}

\caption{The table shows errors in predicted arrival time (column 3rd) estimated using true velocity in COR2 FOV (column 5th) with actual arrival time at L1 (column 2nd). The 4th column shows the CME leading edge velocity measured at L1 point by \textit{in-situ} spacecraft.}
\end{table}

\clearpage

\begin{figure}
\begin{center}
\includegraphics[angle=0,scale=.50]{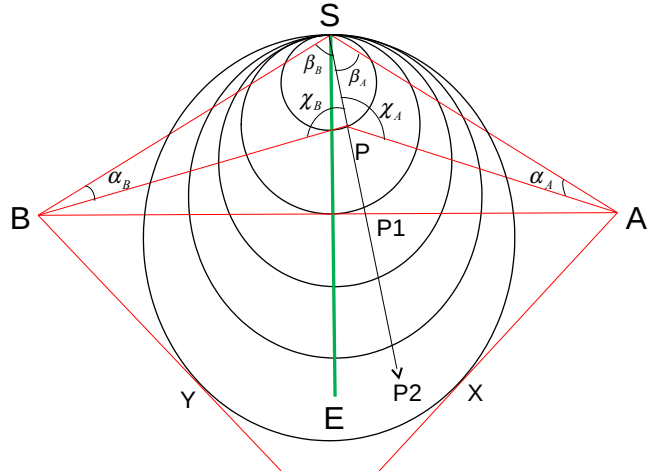}
\caption{\scriptsize{Schematic diagram of geometric triangulation for a moving feature between two spacecraft STEREO A and B, in the direction of arrow. Line SE represents the Sun-Earth line and $\alpha$, $\beta$ and $\chi$ denote the elongation, propagation and scattering angle respectively. Subscript $A$ and $B$ represent angles measured from STEREO A and B view points.}}
\end{center}
\end{figure}

\begin{figure}
\includegraphics[angle=0,scale=.42]{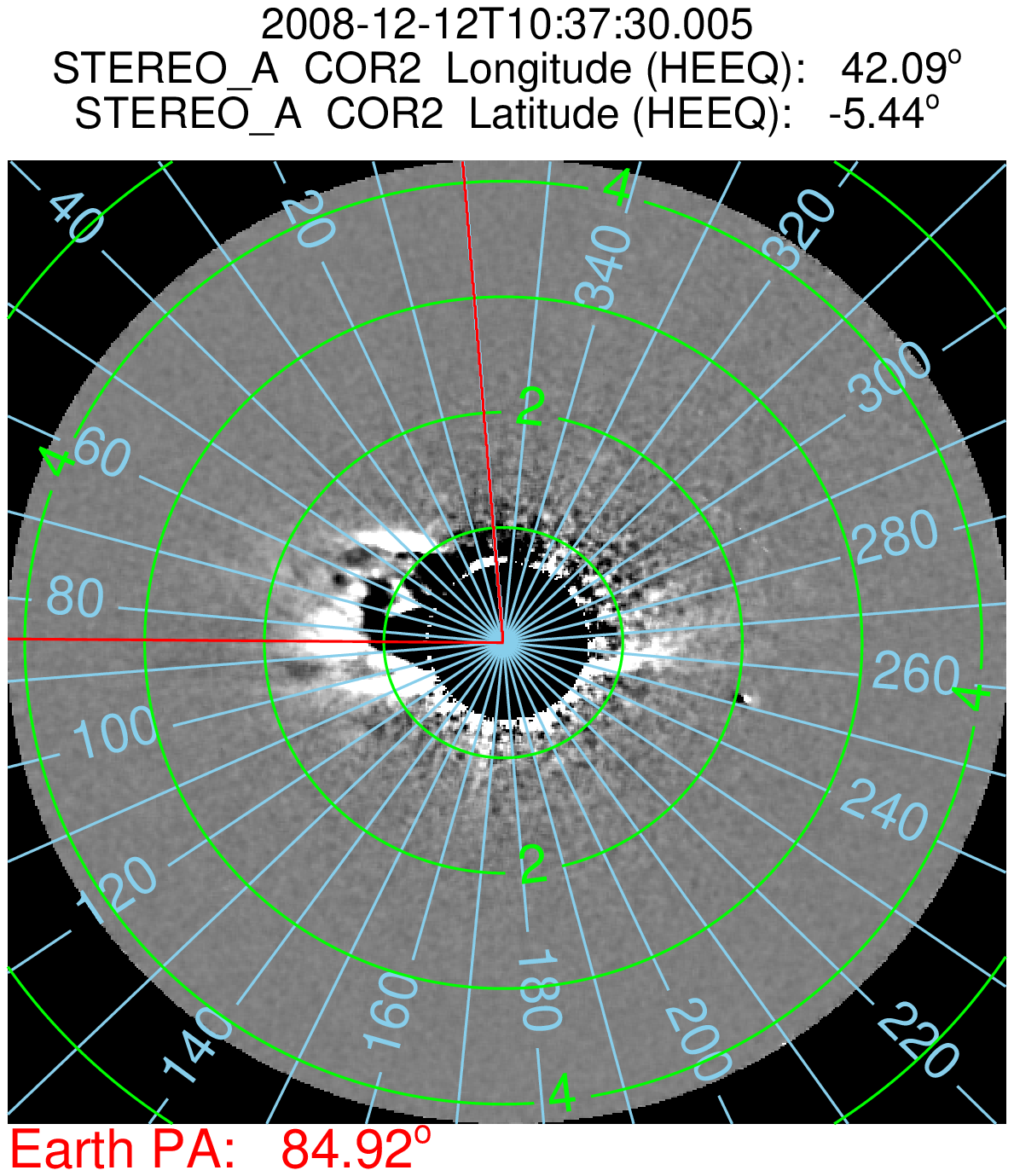}
\includegraphics[angle=0,scale=.42]{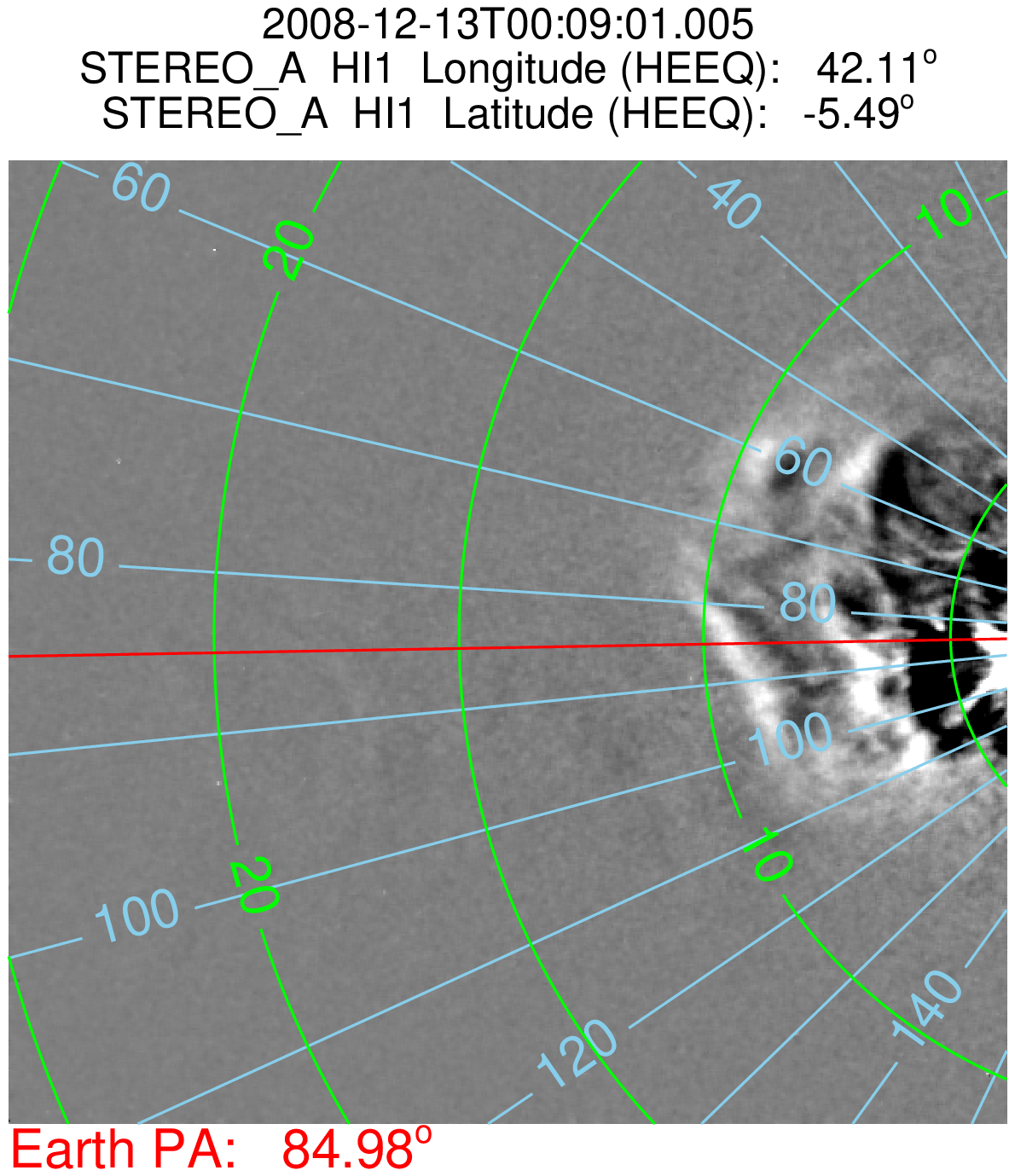}
\includegraphics[angle=0,scale=.42]{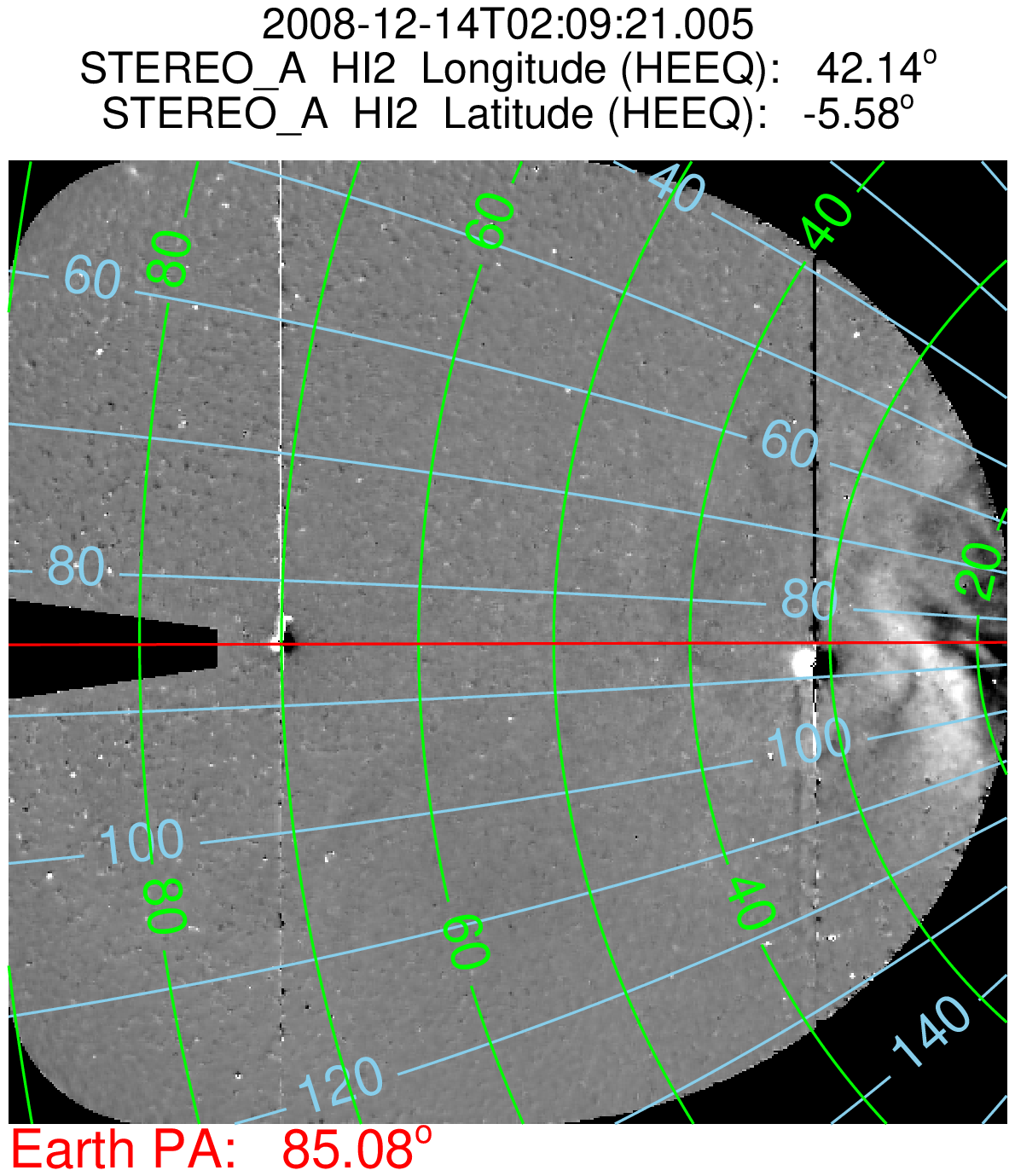}
\caption{\scriptsize{Left, middle and right figures show the running difference image taken by COR2, HI1 and HI2 respectively on STEREO/SECCHI Ahead with contours of elongation angle (green) and position angle (blue) overlaid. The horizontal red line is along the ecliptic at the position angle of Earth. The vertical red line in the left figure marks zero degree position angle.}}
\end{figure}

\clearpage

\begin{figure}
\centering
\includegraphics[angle=0,scale=.50]{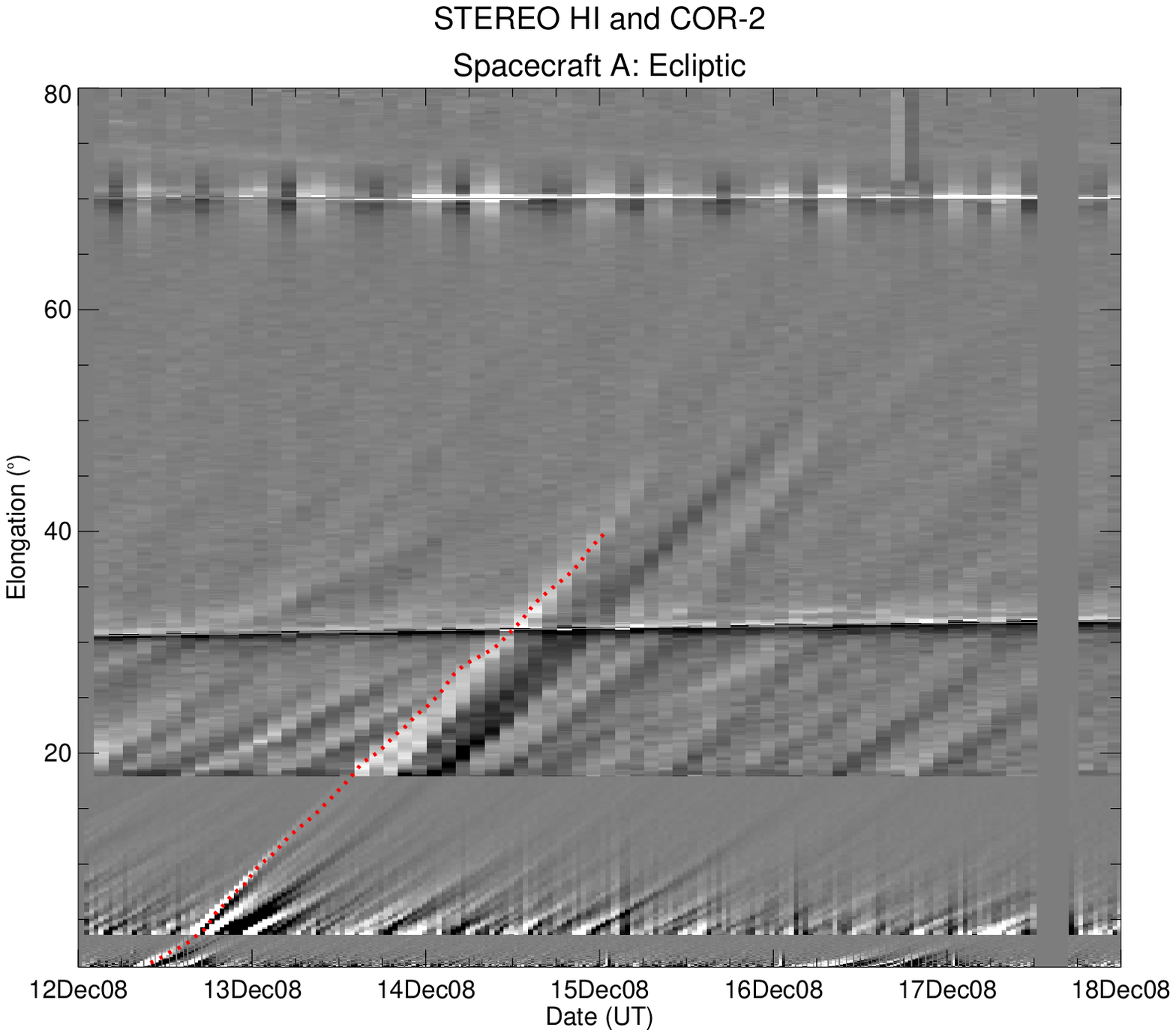}
\includegraphics[angle=0,scale=.50]{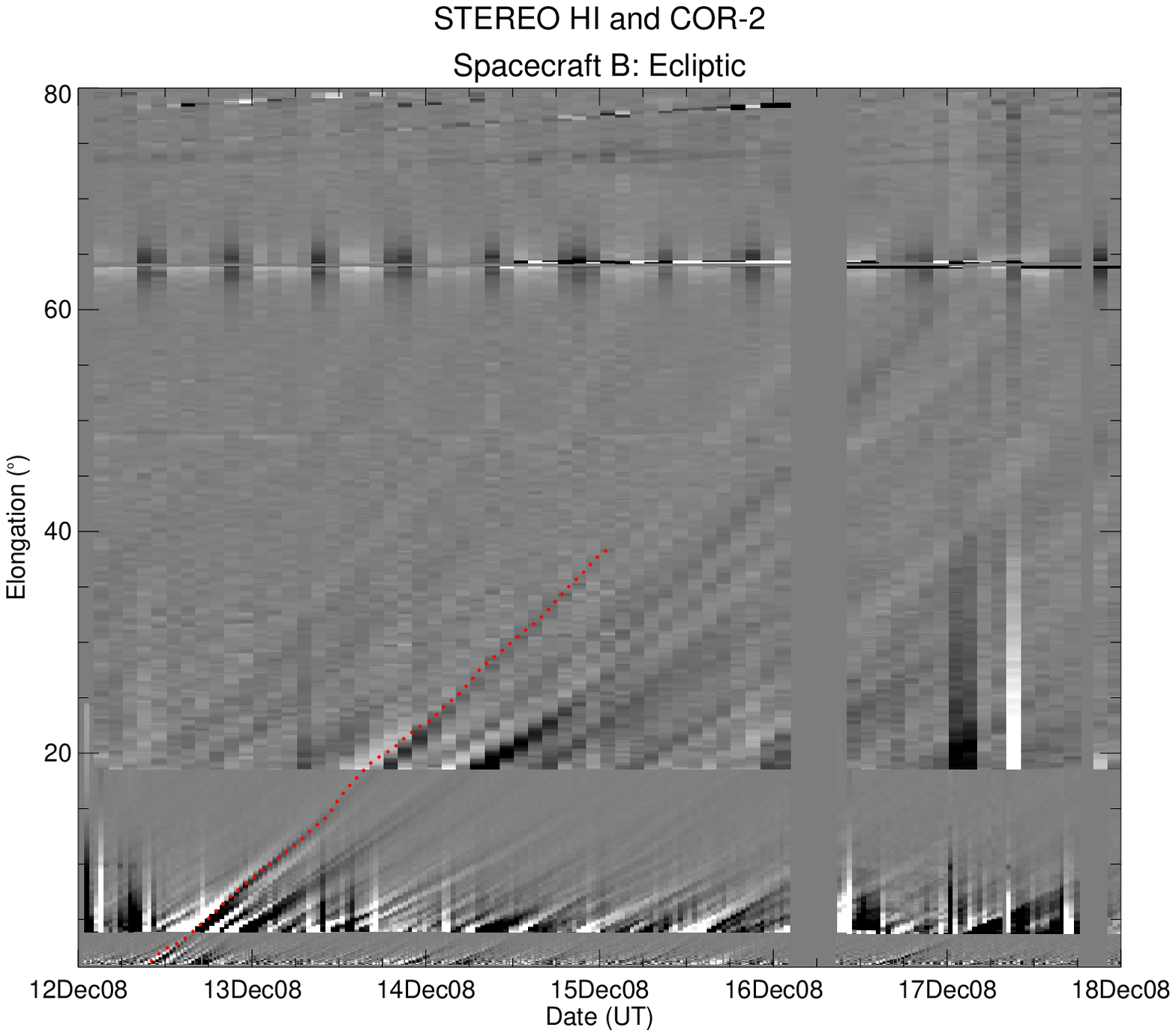}
\caption{\scriptsize{Time elongation map (J-map) for STEREO A (left) and STEREO B (right) for the interval of 12 to 18 December 2008 is shown. The red dot shows the tracked feature corresponding to 12 December 2008 CME. Tracked dot is over plotted on the J-map to mark the elongation variation with time.}}
\end{figure}

\begin{figure}[htbp]
  \begin{minipage}[h]{0.50\linewidth}
    \centering
		\includegraphics[angle=0,scale=0.6]{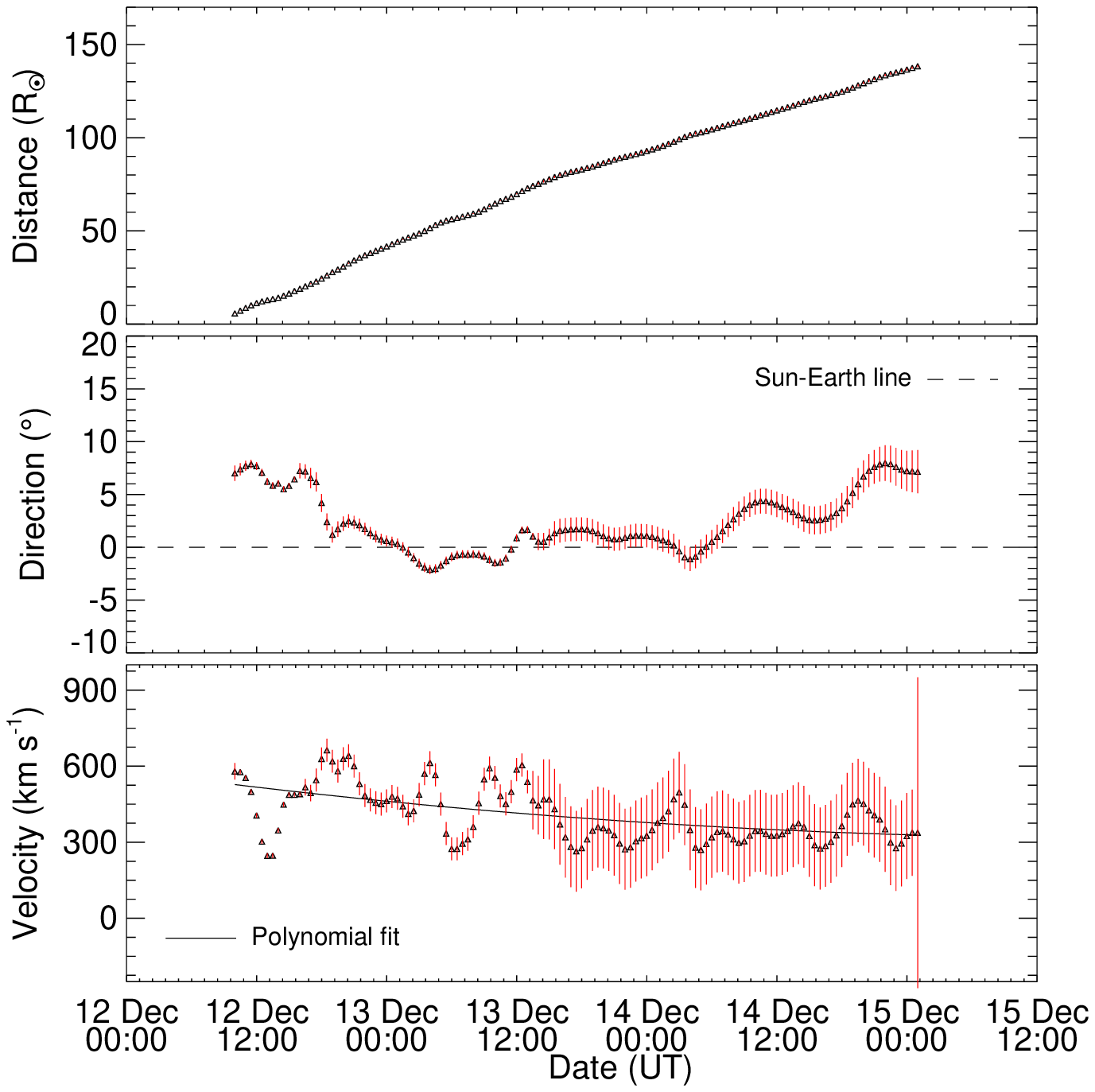}
    \caption{\scriptsize{From top to bottom, panels show the distance, propagation direction and velocity respectively of 12 December 2008 CME. In second panel, horizontal dashed line represents the Sun-Earth line. The red vertical lines show the error bars. In bottom panel velocity is estimated from adjacent distances using three point Lagrange interpolation, therefore have large error bars. The solid line in the third panel is the polynomial fit of actual velocity data points.}}
  \end{minipage}
  \hspace{0.2cm}
  \begin{minipage}[h]{0.45\linewidth}
    \centering
    \includegraphics[scale=0.65]{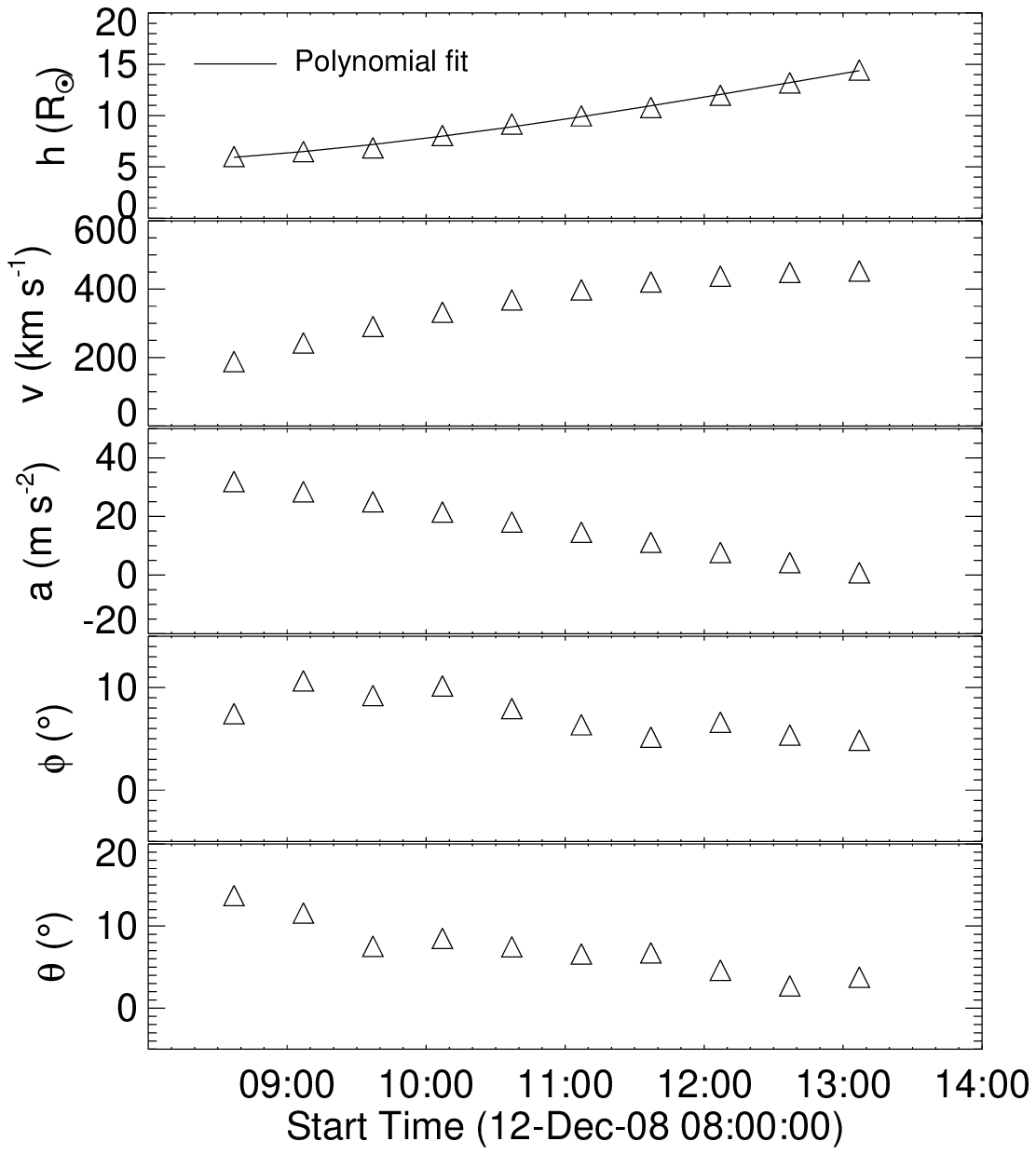}
    \caption{\scriptsize{From top to bottom, the panels show the estimated true height, radial velocity, acceleration, longitude and latitude and time on X- axis using tie-pointing method. Velocity and acceleration is calculated by first and second order differentiation respectively, by fitting third order polynomial to true height data.}}
  \end{minipage}
\end{figure}

\begin{figure}[htbp]
  \begin{minipage}[h]{0.45\linewidth}
    \centering
		\includegraphics[angle=0,scale=0.45]{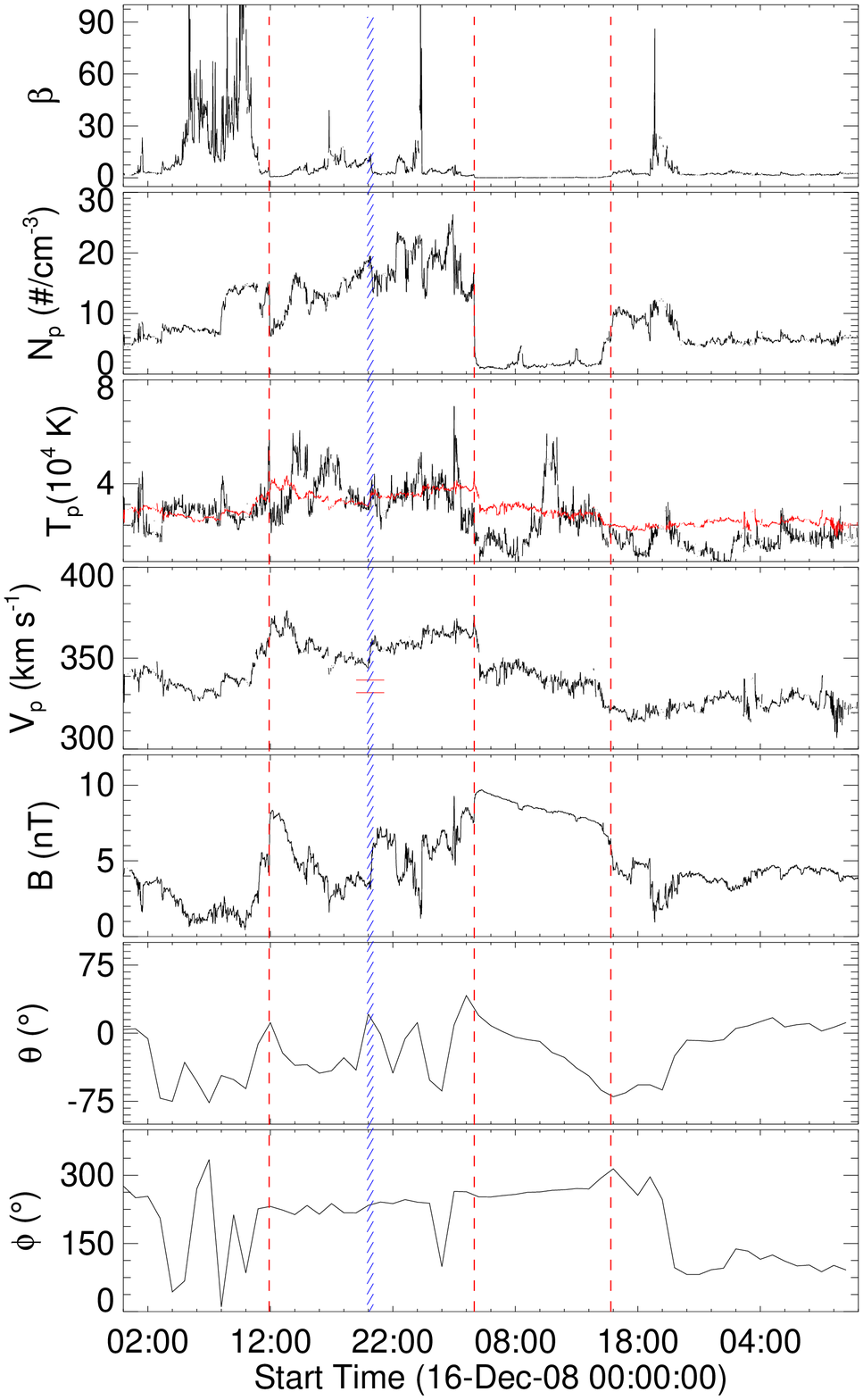}
    \caption{\scriptsize{From top to bottom, panels show the variation of plasma beta, proton density, proton temperature, proton velocity, magnitude of magnetic field, latitude and longitude of magnetic field vector respectively corresponding to CME of 12 December 2008. From the left, the first, second and third vertical dashed lines (red) mark the arrival time of CME sheath, leading and trailing edge of magnetic cloud, respectively. The hatched region (blue) marks the interval of predicted arrival time of tracked feature. From the top, in the third panel the expected proton temperature is shown as red curve and in the fourth panel horizontal lines (red) mark the predicted velocities of tracked feature at L1.}}
    
  \end{minipage}
  \hspace{0.2cm}
  \begin{minipage}[h]{0.50\linewidth}
    \centering
    \includegraphics[scale=0.54]{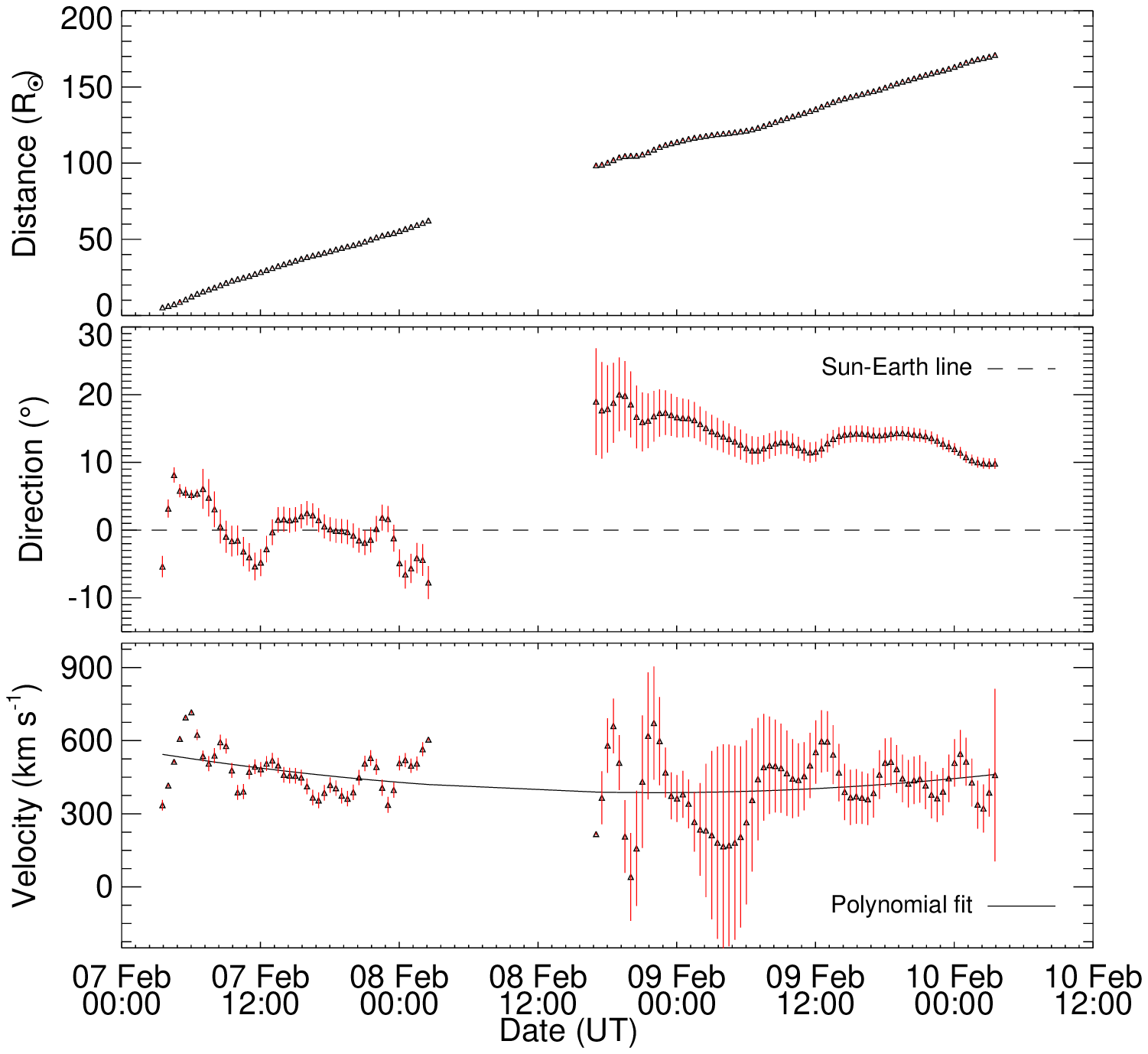}
    \caption{\scriptsize{From top to bottom, panels show the distance, propagation direction and velocity respectively of 07 February 2010 event. The details are mentioned in Figure 4 caption.}}
    
  \end{minipage}
\end{figure}

\begin{figure}[htbp]
  \begin{minipage}[h]{0.45\linewidth}
    \centering
		\includegraphics[angle=0,scale=0.42]{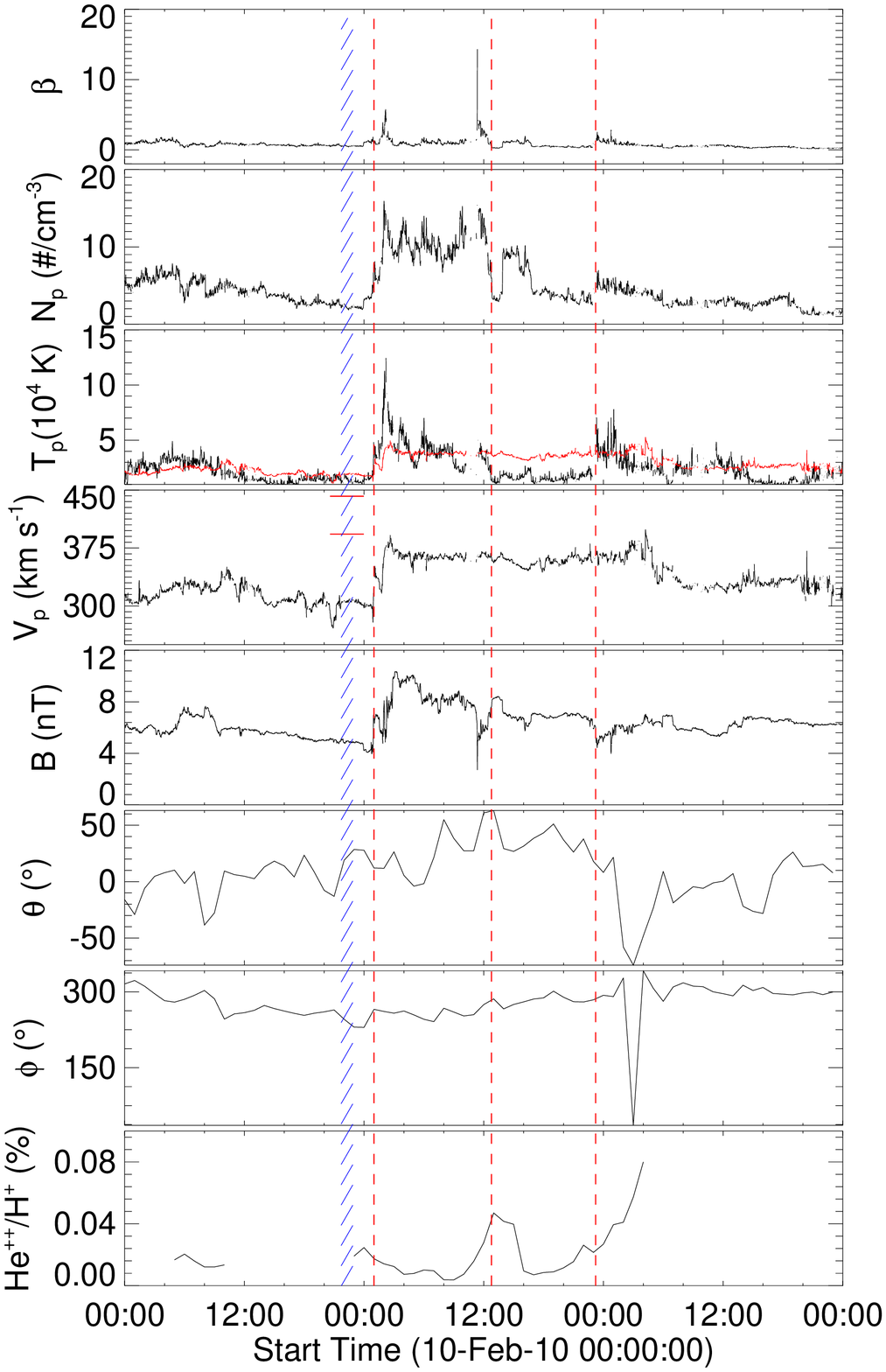}
    \caption{\scriptsize{From top to bottom, panels show the plasma beta, proton density, proton temperature, proton velocity, magnitude of magnetic field, latitude, longitude of magnetic field vector and alpha to proton ratio respectively corresponding to CME of 07 February 2010. From the left, the first, second and third vertical dashed lines (red) mark the arrival time of sheath, leading and trailing edge of CME, respectively. The hatched region (blue) shows the interval of predicted arrival time of tracked feature. From the top, in the third panel the expected proton temperature is shown as red curve and in the fourth panel two horizontal lines (red) mark the predicted velocities of tracked feature at L1.}}
    
  \end{minipage}
  \hspace{0.2cm}
  \begin{minipage}[h]{0.50\linewidth}
    \centering
    \includegraphics[scale=0.45]{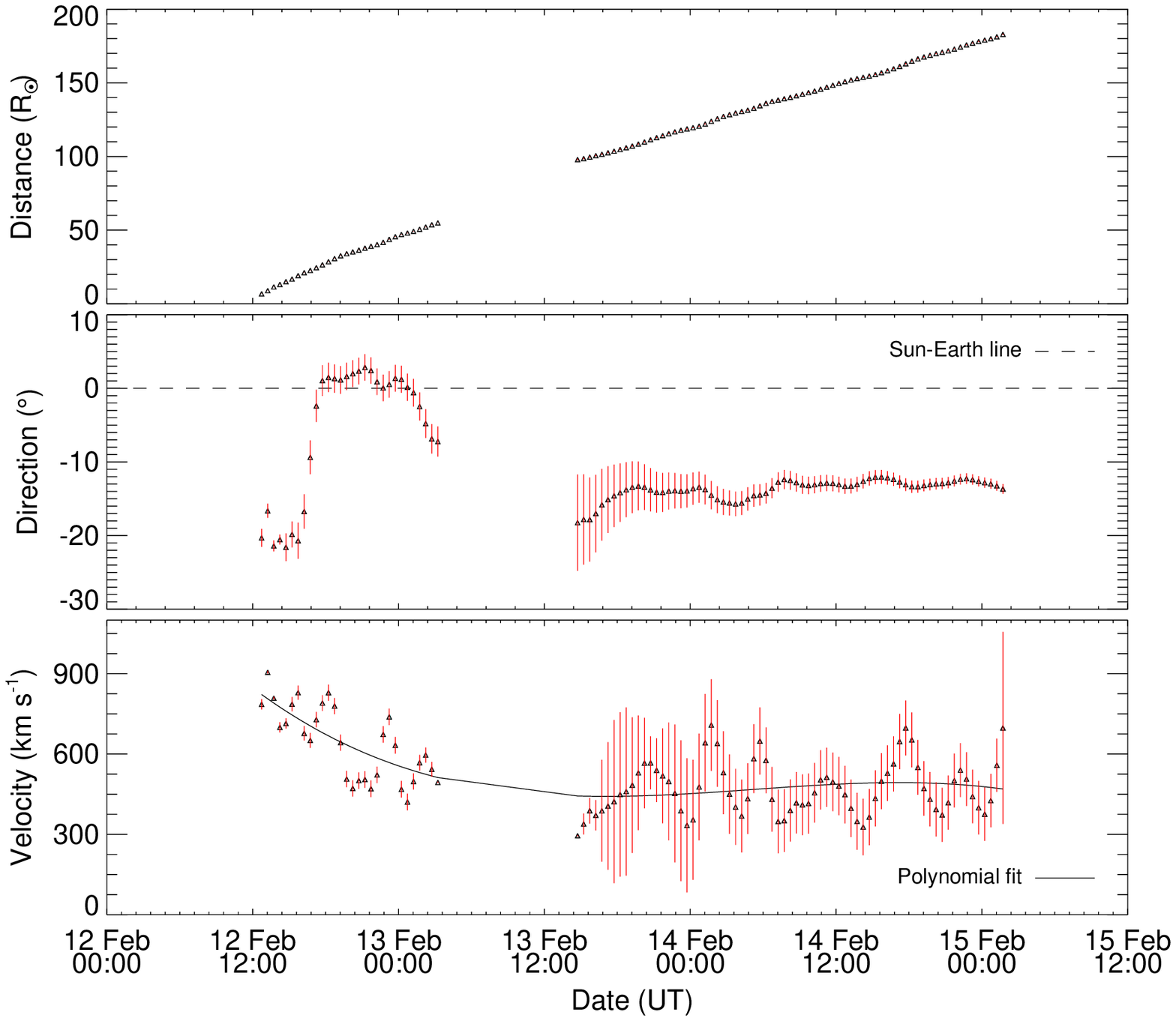}
    \caption{\scriptsize{From top to bottom, panels show the distance, propagation direction and velocity respectively of 12 February 2010 event. For details see Figure 4 caption.}}
    
  \end{minipage}
\end{figure}

\begin{figure}[htbp]
  \begin{minipage}[h]{0.45\linewidth}
    \centering
		\includegraphics[angle=0,scale=0.42]{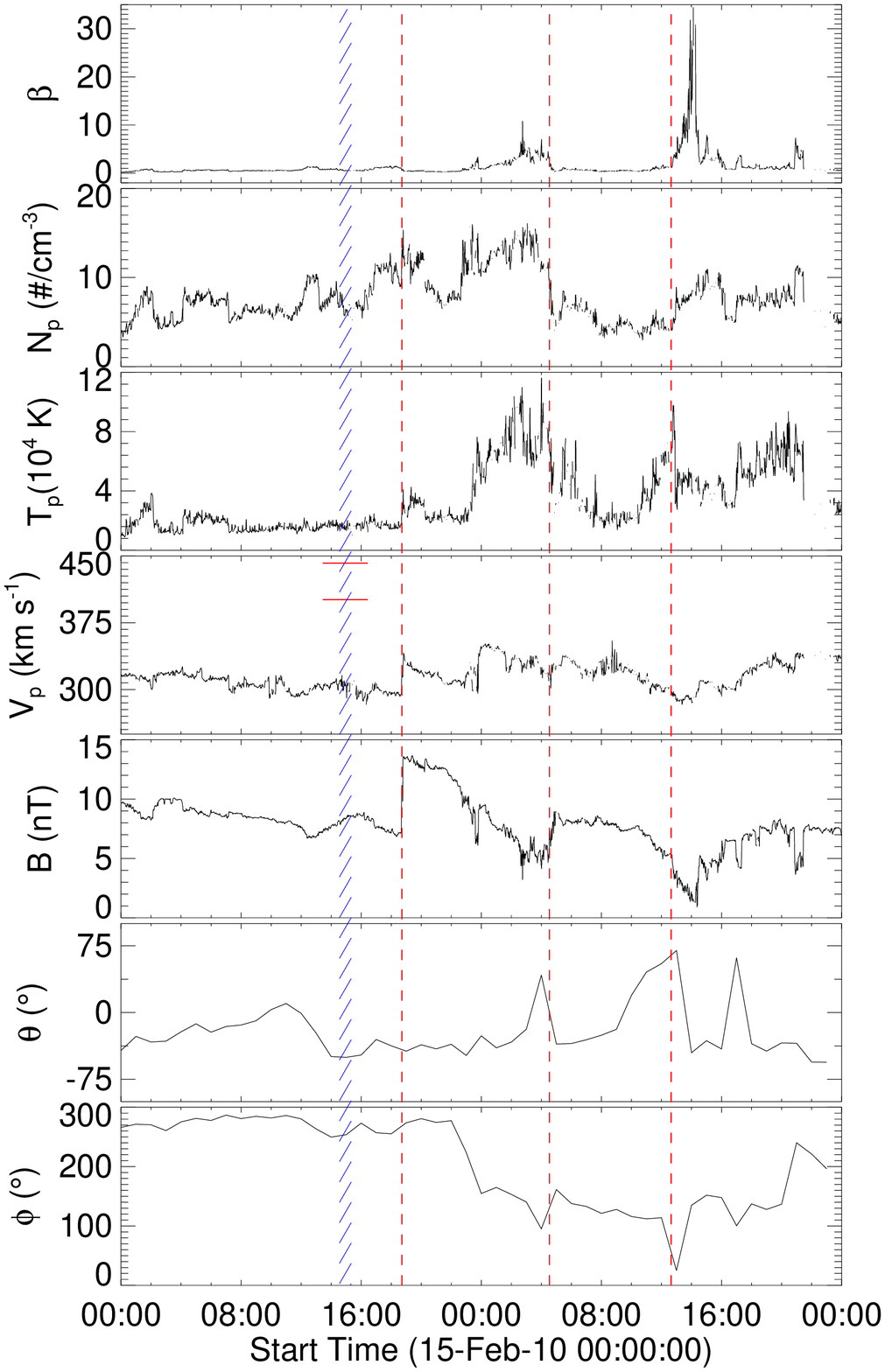}
    \caption{\scriptsize{From top to bottom, panels show the plasma beta, proton density, proton temperature, proton velocity, magnitude of magnetic field, latitude and longitude of magnetic field vector respectively corresponding to CME of 12 February 2010. From the left, the first, second and third vertical dashed lines (red) mark the arrival time of shock, leading and trailing edge of CME, respectively. The hatched region (blue) marks the interval of predicted arrival time of tracked feature. In the fourth panel, horizontal lines (red) mark the predicted velocities of tracked feature at L1.}}
    
  \end{minipage}
  \hspace{0.2cm}
  \begin{minipage}[h]{0.50\linewidth}
    \centering
    \includegraphics[scale=0.52]{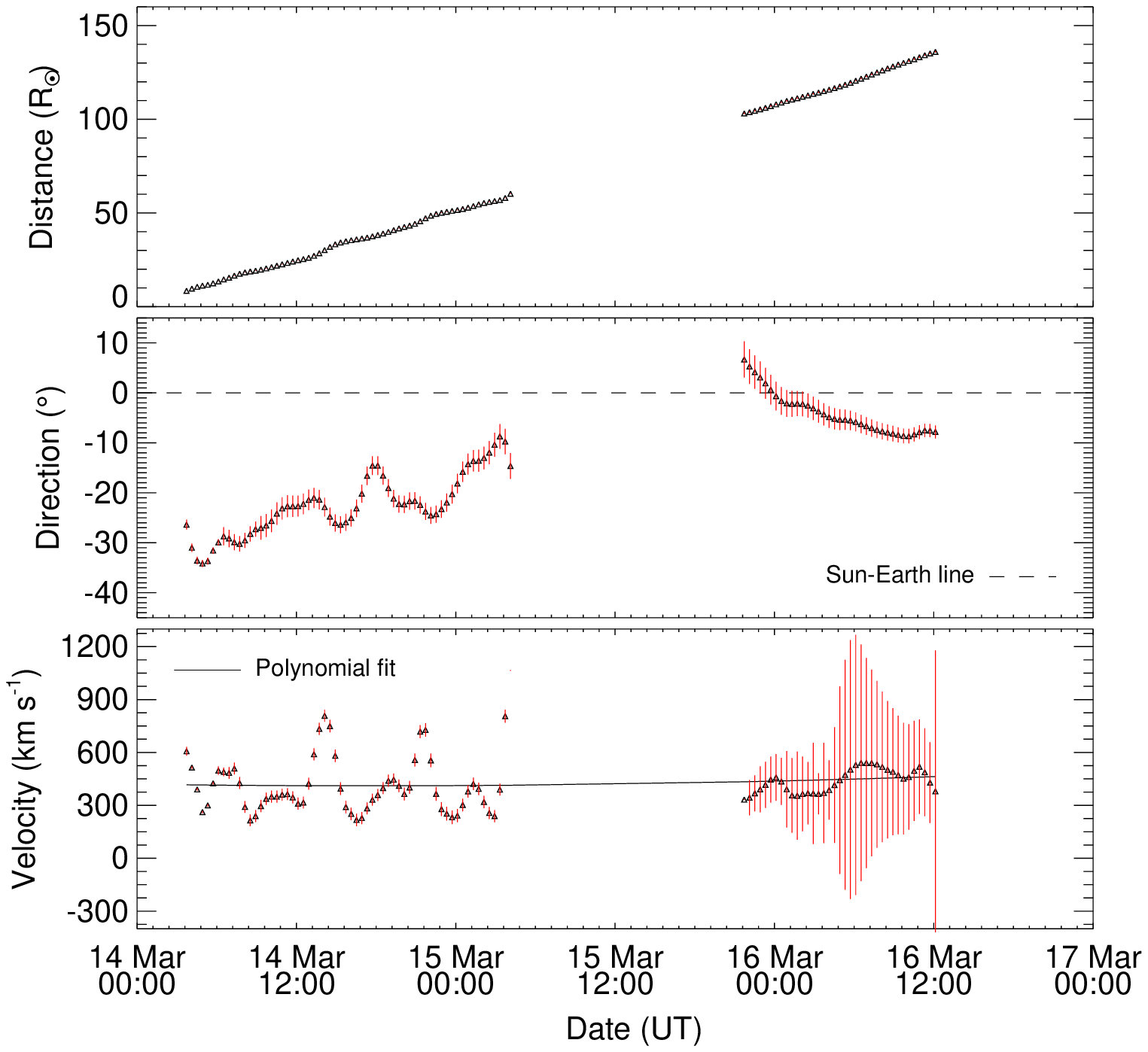}
    \caption{\scriptsize{From top to bottom, panels show the distance, propagation direction and velocity respectively of 14 March  2010 event. For details see Figure 4 caption.}}
    
  \end{minipage}
\end{figure}

\begin{figure}[htbp]
  \begin{minipage}[h]{0.45\linewidth}
    \centering
		\includegraphics[angle=0,scale=0.42]{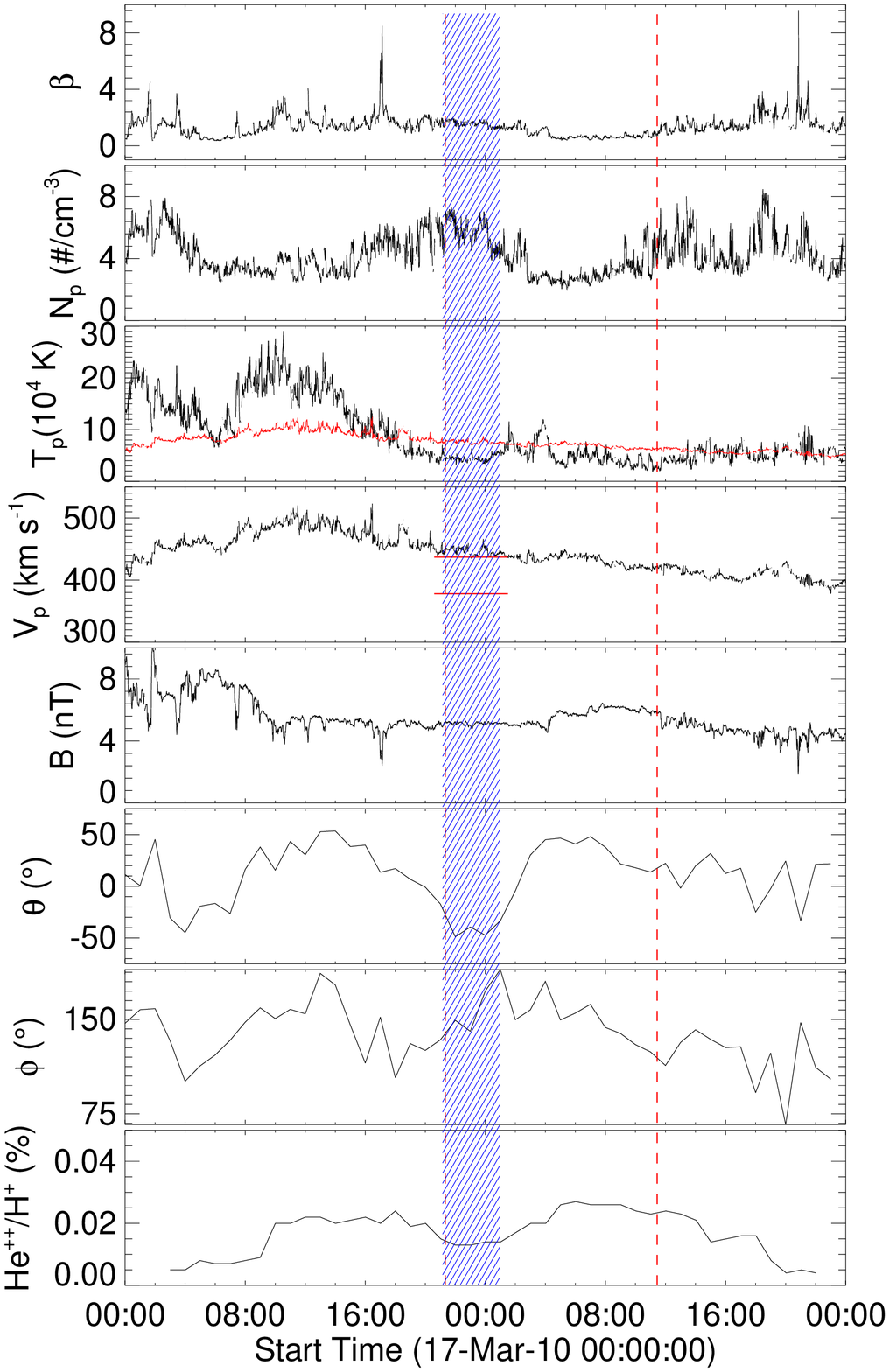}
    \caption{\scriptsize{From top to bottom, panels show the plasma beta, proton density, proton temperature, proton velocity, magnitude of magnetic field, latitude, longitude of magnetic field vector and alpha to proton ratio respectively corresponding to CME of 14 March 2010. From the left, the first and second vertical dashed lines (red) mark the arrival time of CME leading and trailing edge, respectively. The hatched region (blue) marks the interval of predicted arrival time of tracked feature. From the top, in the third panel the expected proton temperature is shown as red curve and in the fourth panel two horizontal lines (red) mark the predicted velocities of tracked feature at L1.}}
    
  \end{minipage}
  \hspace{0.2cm}
  \begin{minipage}[h]{0.50\linewidth}
    \centering
    \includegraphics[scale=0.58]{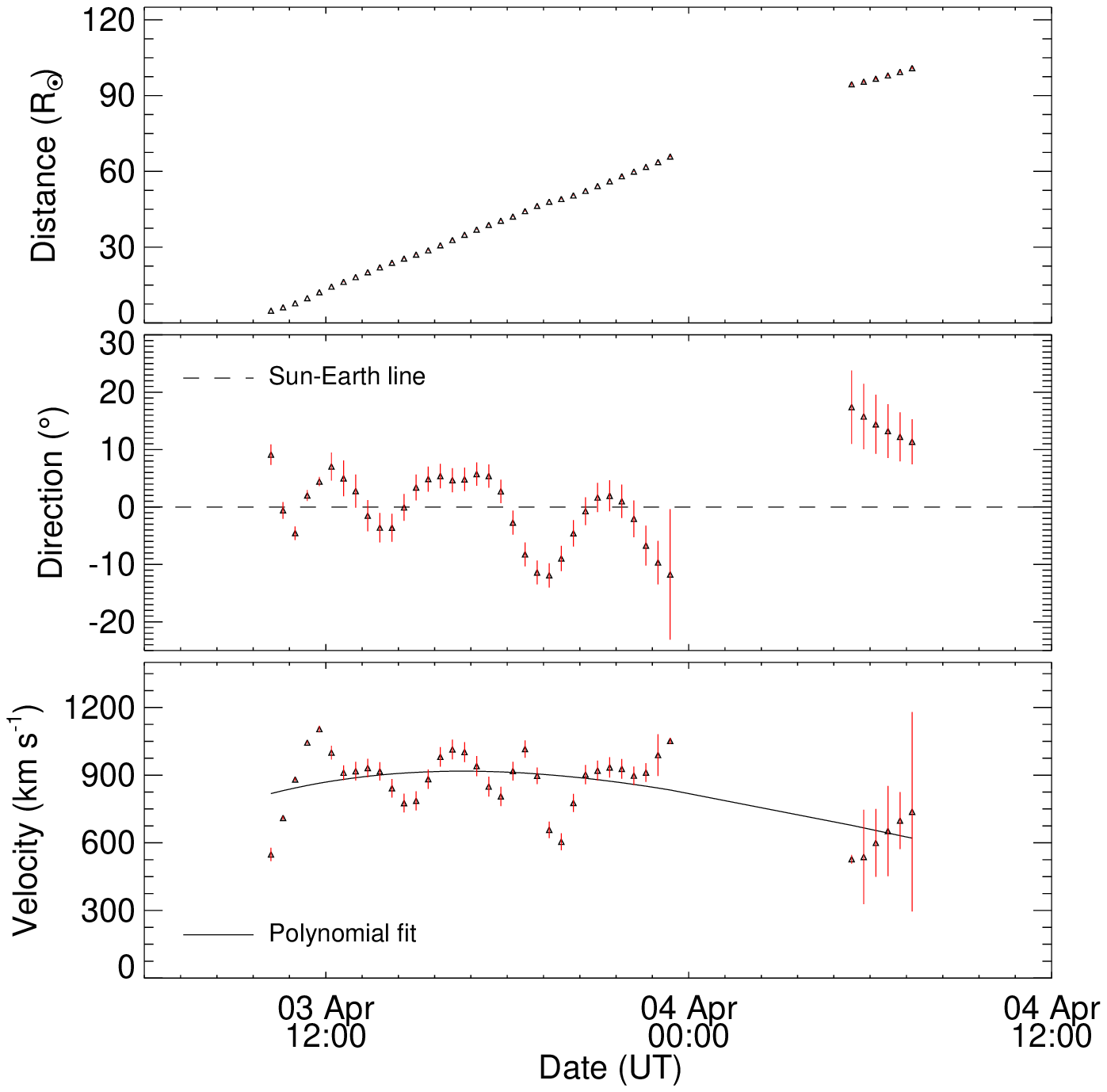}
    \caption{\scriptsize{From top to bottom, panels show the distance, propagation direction and velocity respectively of 03 April 2010 event. For details see Figure 4 caption.}}
    
  \end{minipage}
\end{figure}

\begin{figure}[htbp]
  \begin{minipage}[h]{0.45\linewidth}
    \centering
		\includegraphics[angle=0,scale=0.43]{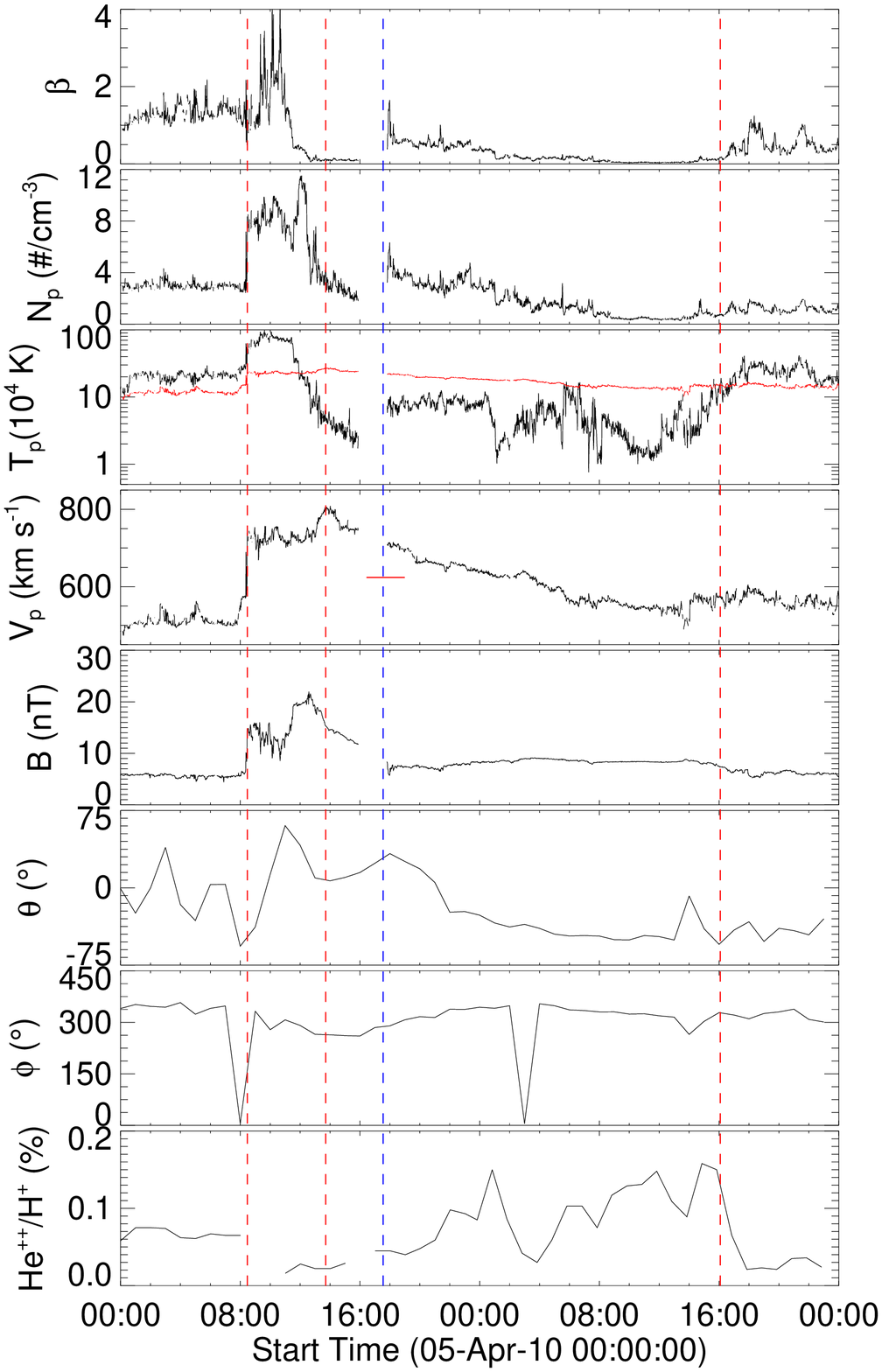}
    \caption{\scriptsize{From top to bottom, panels show the plasma beta, proton density, proton temperature, proton velocity, magnitude of magnetic field, latitude, longitude of magnetic field vector and alpha to proton ratio respectively corresponding to CME of 03 April 2010. From the left, the first, second and fourth vertical dashed lines (red) mark the arrival time of shock ,leading and trailing edge of CME, respectively. The third vertical dashed line (blue) marks the predicted arrival time of tracked feature. From the top, in the third panel the expected proton temperature is shown as red curve and in the fourth panel horizontal line (red) marks the predicted velocities of tracked feature at L1.}}
    
  \end{minipage}
  \hspace{0.2cm}
  \begin{minipage}[h]{0.50\linewidth}
    \centering
    \includegraphics[scale=0.58]{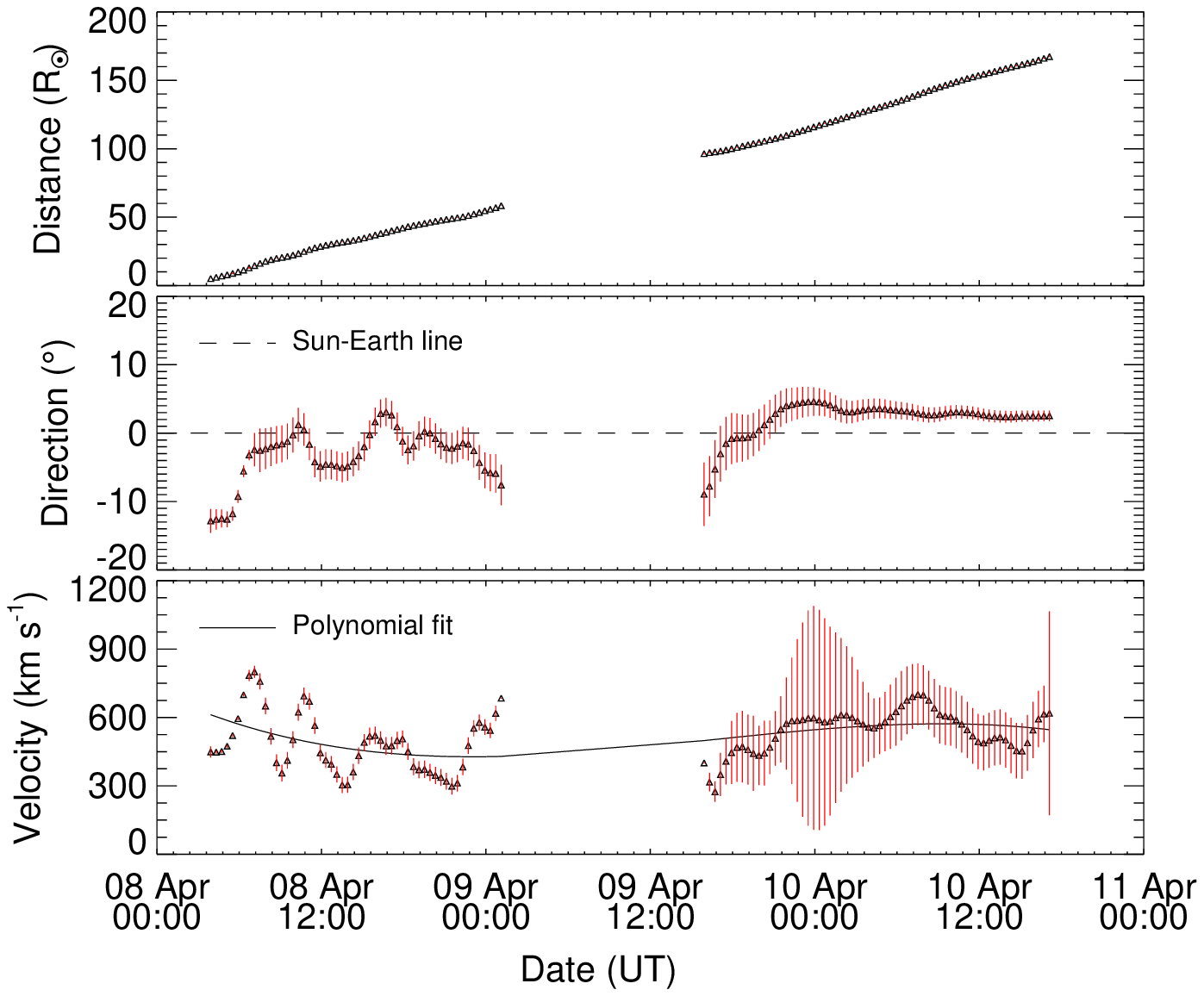}
    \caption{\scriptsize{From top to bottom, panels show the distance, propagation direction and velocity respectively of 08 April  2010 event. For details see Figure 4 caption.}}
    
  \end{minipage}
\end{figure}

\begin{figure}[htbp]
  \begin{minipage}[h]{0.45\linewidth}
    \centering
		\includegraphics[angle=0,scale=0.42]{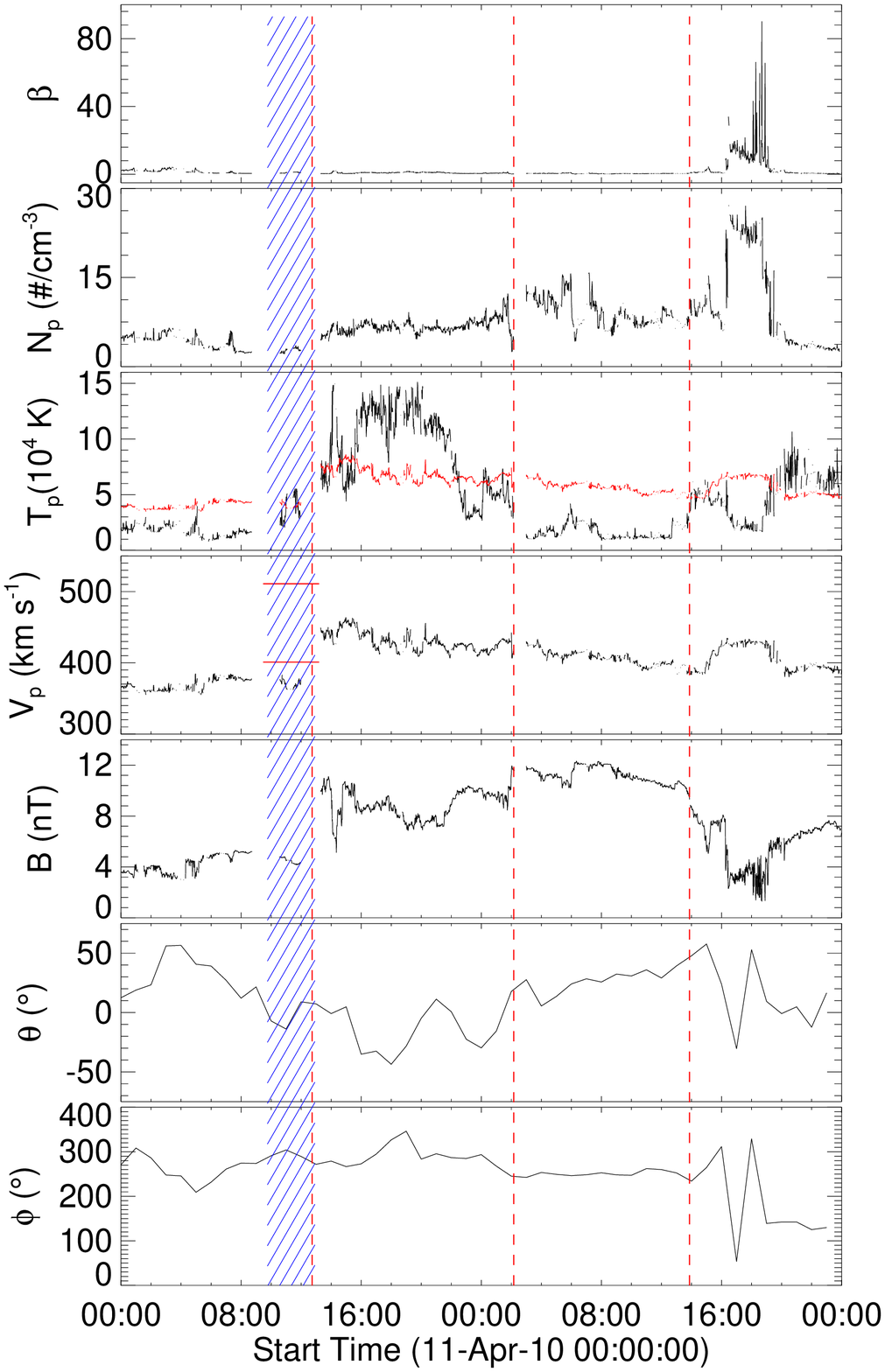}
    \caption{\scriptsize{From top to bottom, panels show the plasma beta, proton density, proton temperature, proton velocity, magnitude of magnetic field, latitude and longitude of magnetic field vector respectively corresponding to CME of 08 April 2010. From the left, the first, second and third  vertical dashed lines (red) mark the arrival time of sheath, leading and trailing edge of magnetic cloud, respectively. The hatched region (blue) marks the interval of predicted arrival time of tracked feature. From the top, in the third panel the expected proton temperature is shown as red curve and in the fourth panel two horizontal lines (red) mark the predicted velocities of tracked feature at L1.}}
    
  \end{minipage}
  \hspace{0.2cm}
  \begin{minipage}[h]{0.50\linewidth}
    \centering
    \includegraphics[scale=0.64]{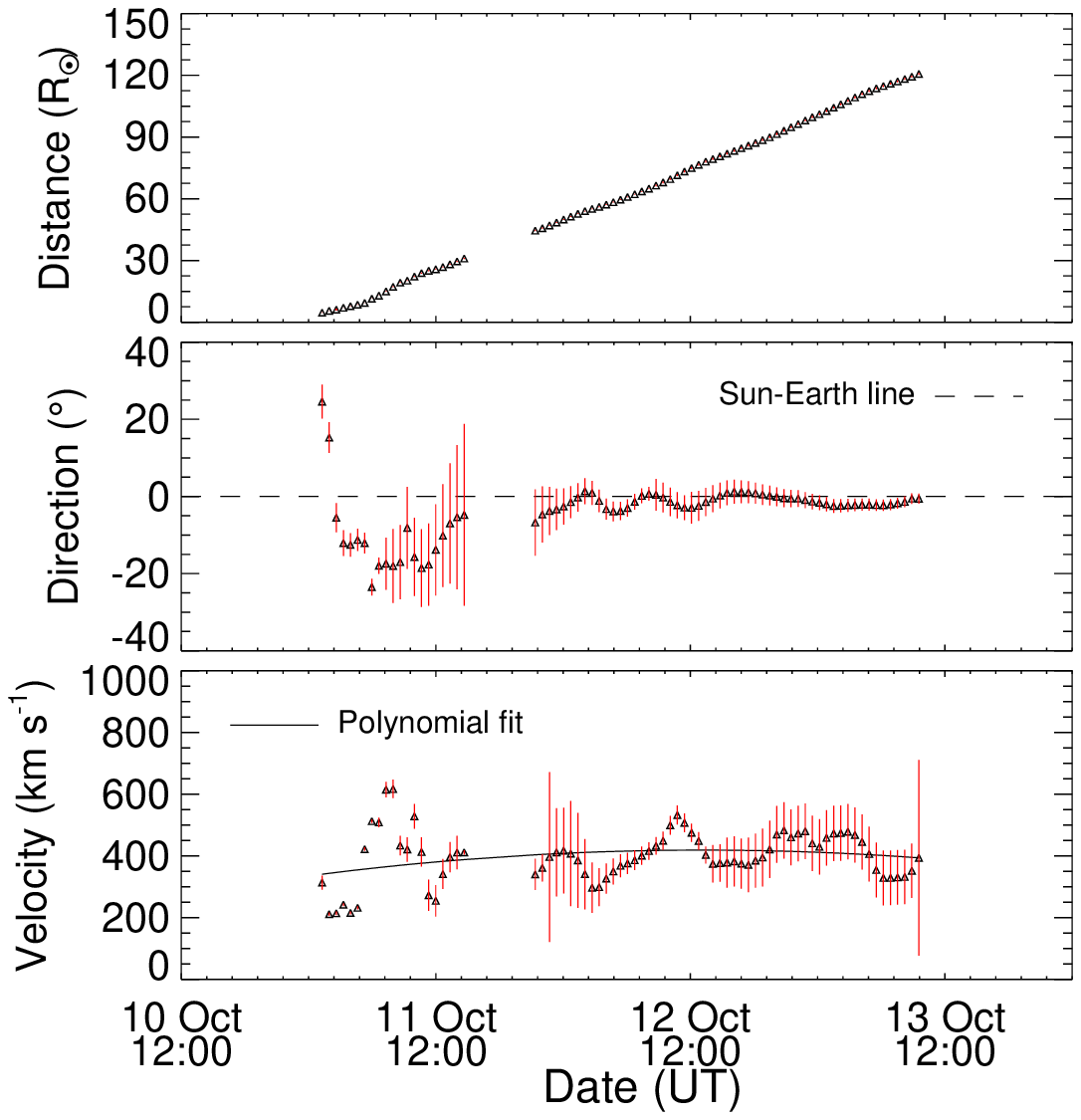}
    \caption{\scriptsize{From top to bottom, panels show the distance, propagation direction and velocity respectively of 10 October 2010. For details see Figure 4 caption.}}
    
  \end{minipage}
\end{figure}

\begin{figure}[htbp]
  \begin{minipage}[h]{0.45\linewidth}
    \centering
		\includegraphics[angle=0,scale=0.43]{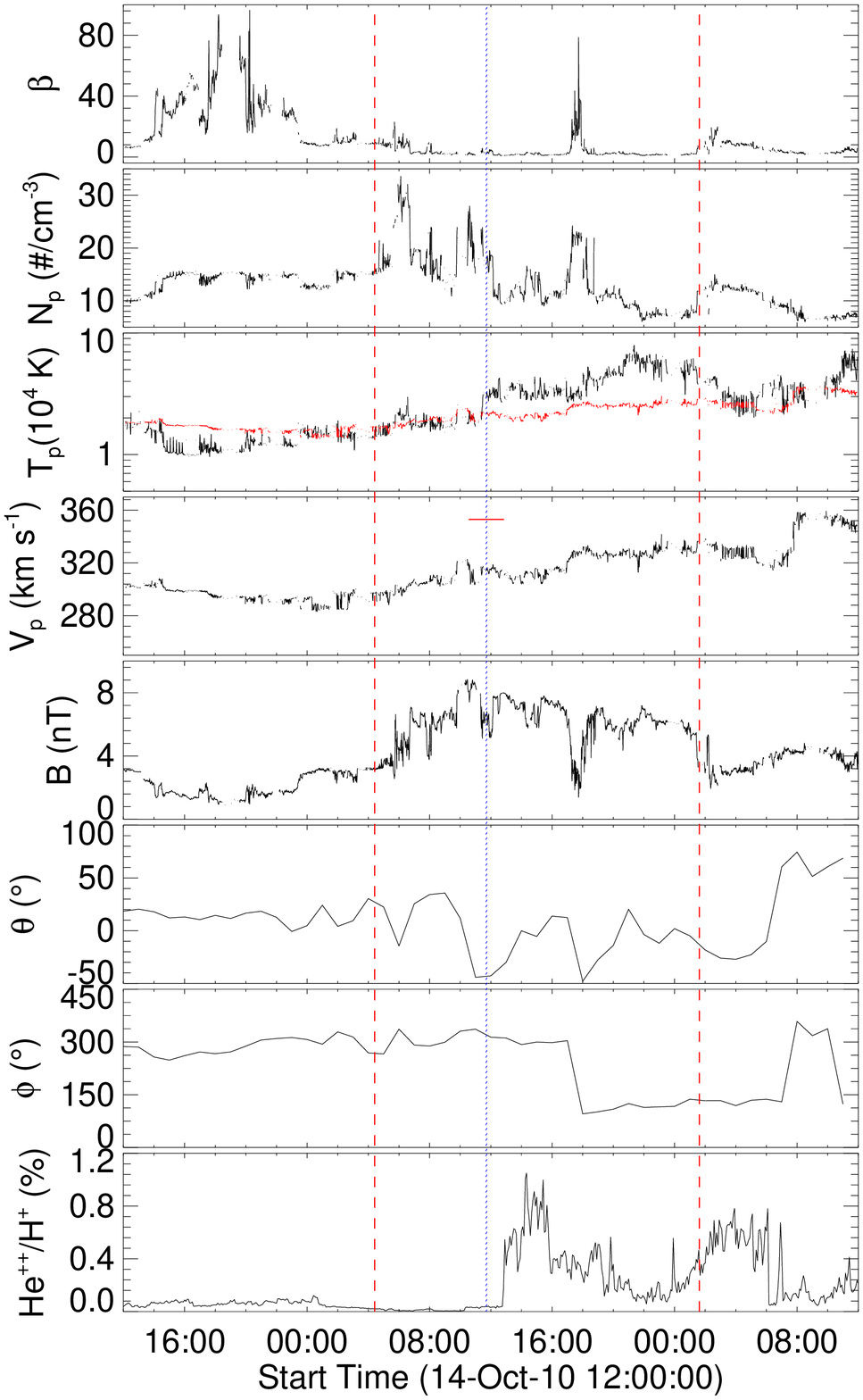}
    \caption{\scriptsize{From top to bottom, panels show the plasma beta, proton density, proton temperature, proton velocity, magnitude of magnetic field, latitude, longitude of magnetic field vector and alpha to proton ratio respectively corresponding to CME of 10 October 2010. From the left, the first and second vertical dashed lines (red) mark the arrival time of leading and trailing edge of CME sheath, respectively. The hatched line (blue) marks the interval of predicted arrival time of tracked feature. From the top, in the third panel the expected proton temperature is shown as red curve and in the fourth panel two horizontal lines (red) mark the predicted velocities of tracked feature at L1.}}
    
  \end{minipage}
  \hspace{0.2cm}
  \begin{minipage}[h]{0.50\linewidth}
    \centering
    \includegraphics[scale=0.65]{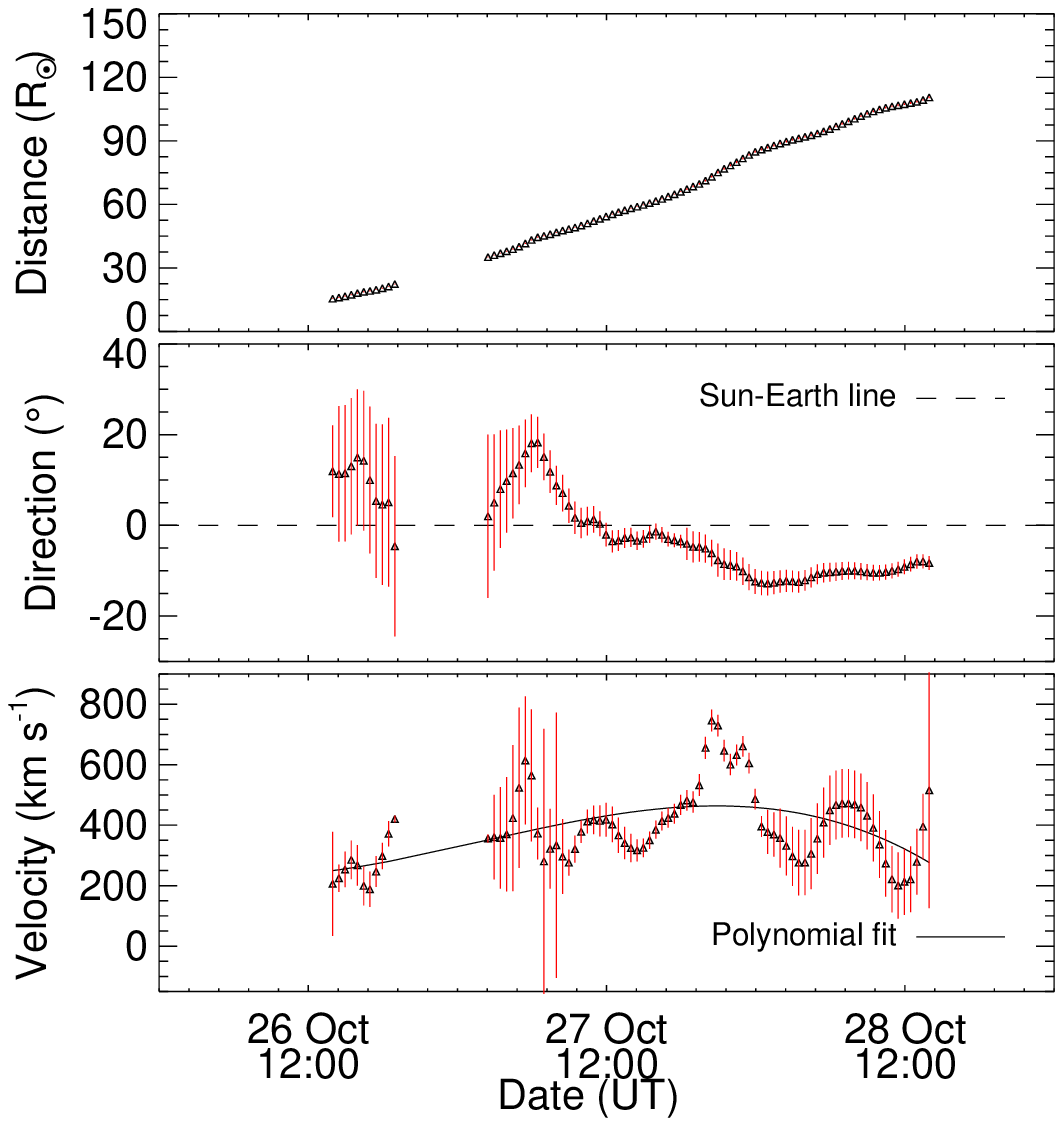}
    \caption{\scriptsize{From top to bottom, panels show the distance, propagation direction and velocity respectively of 26 October 2010. For details see Figure 4 caption.}}
    
  \end{minipage}
\end{figure}

\begin{figure}
\begin{center}
\includegraphics[angle=0,scale=0.50]{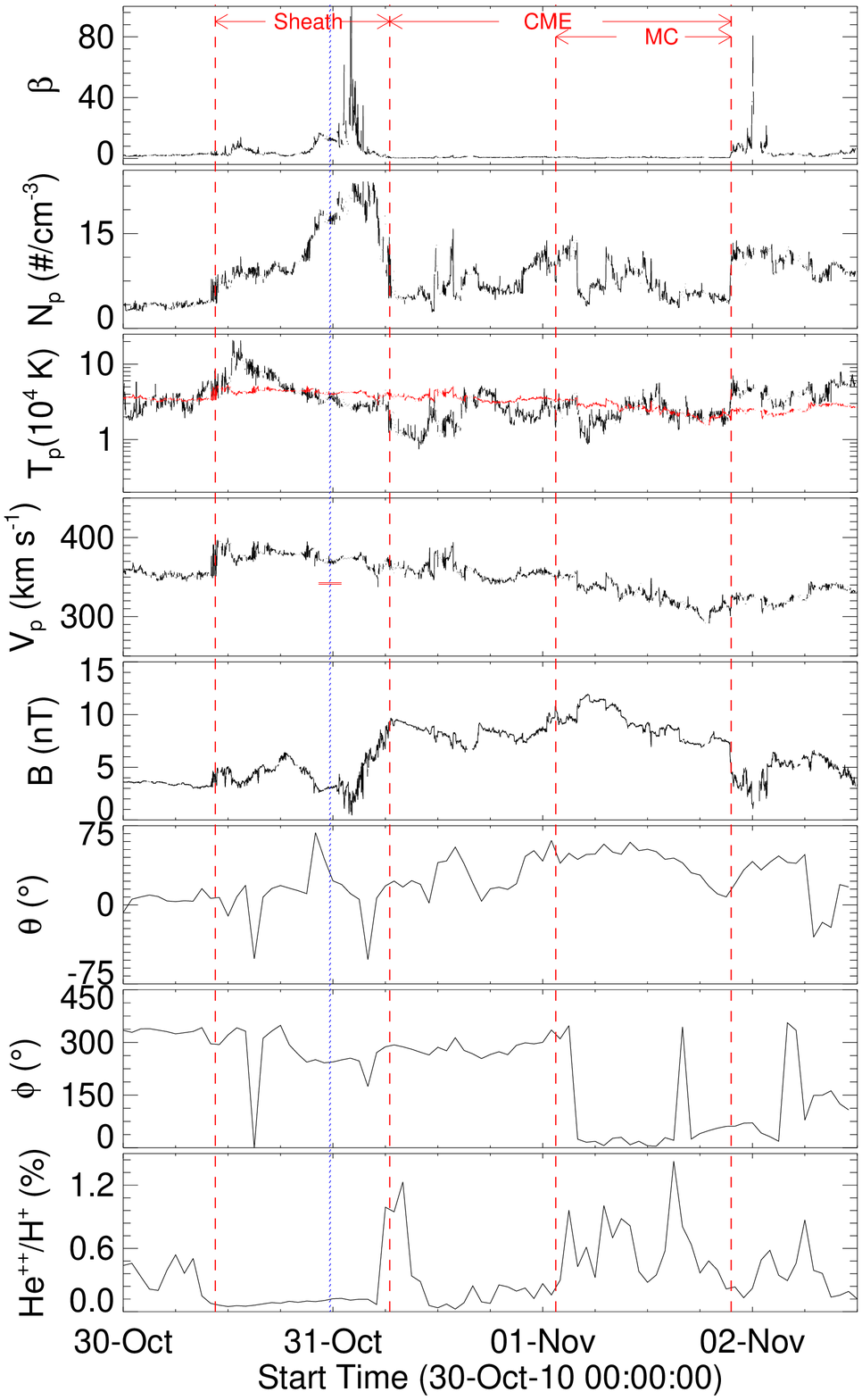}
\caption{\scriptsize{From top to bottom, panels show the plasma beta, proton density, proton temperature, proton velocity, magnitude of magnetic field, latitude, longitude of magnetic field vector and alpha to proton ratio respectively corresponding to CME of 26 October 2010. From the left, the first, second, third and fourth vertical dashed lines (red) mark the arrival time of shock, trailing edge of CME sheath, leading edge and trailing edge of magnetic cloud, respectively. The hatched line (blue) marks the interval of predicted arrival time of tracked feature. From the top, in the third panel the expected proton temperature is shown as red curve and in the fourth panel two horizontal lines (red) mark the predicted velocities of tracked feature at L1.}}
\end{center}
\end{figure}

\end{document}